\documentclass[aps,prl,twocolumn,groupedaddress,showpacs]{revtex4-1}
\pdfoutput=1
\usepackage[utf8]{inputenc}
\usepackage[english]{babel}
\usepackage{amsmath,graphicx,enumerate}

\usepackage{color}

\begin{document}

\title{Thermal convection in granular gases with dissipative lateral walls} 
\author{Giorgio Pontuale$^{1}$}
\author{Andrea Gnoli$^{1}$}
\author{Francisco Vega Reyes$^{1,2}$}
\author{Andrea Puglisi$^{1}$} 
\affiliation{$^1$Istituto dei Sistemi Complessi - CNR and Dipartimento di Fisica, Universit\`a di Roma Sapienza, P.le Aldo Moro 2, 00185, Rome, Italy \\
$^2$Departamento de F\'isica and Instituto de Computaci\'on Cient\'ifica Avanzada (ICCAEx), Universidad de Extremadura, 06071 Badajoz, Spain}

\date{\today}

\begin{abstract}
We consider a granular gas under the action of gravity, fluidized by a
vibrating base. We show that a horizontal temperature gradient, here
induced by limiting dissipative lateral walls (DLW), leads always to a
granular thermal convection (DLW-TC) that is essentially different
from ordinary bulk-buoyancy-driven convection (BBD-TC). In an experiment
where BBD-TC is inhibited, by reducing gravity with an inclined plane,
we always observe a DLW-TC cell next to each lateral wall. Such a cell
squeezes towards the nearest wall as the gravity and/or the number of
grains increase. Molecular dynamics simulations reproduce the
experimental results and indicate that at large gravity or number of
grains the DLW-TC is barely detectable.
\end{abstract}


\maketitle

Shaken granular media escape most of the laws of equilibrium
thermodynamics and statistical mechanics~\cite{JNB96b}, ranging from
slow plastic flows~\cite{andreotti13} to fast gas-like
dynamics~\cite{D01,poeschel,puglio15}. In the wide granular
phenomenology~\cite{AT06}, an ubiquitous pattern is the convective
cell. Notwithstanding its widespread occurrence, many different
mechanisms lead to granular convection, and their relevance depends on
the granular state under scrutiny.

At high packing fraction and low fluidization, ``dense convection'' is
observed~\cite{laroche89,knight93,ehrichs95,knight96}. A convincing
explanation of dense convection comes from the asymmetric tangential
friction at the lateral walls that results in a net downward shear
force~\cite{gallas92,knight97,grossman97b}.  Dense granular convection
can also involve more complex mechanisms, including the formation of
unstable heaps at the free surface~\cite{aoki96}.

In highly fluidized states (granular gases), the only known
theoretical mechanism is bulk buoyancy-driven thermal convection
(BBD-TC), first observed in simulations~\cite{ramirez00,sunthar01}. In
analogy to molecular liquids~\cite{landau71,chandra81}, it originates
from the buoyancy force associated to temperature/density gradients
that, because of the intrinsic bulk inelasticity, emerge
spontaneously~\cite{VU09}, even with an open or reflecting top
boundary~\cite{brey98c}. BBD-TC is an instability of the hydrostatic
state which requires a combination of parameters (including
inelasticity, gravity and dimensions) to overcome a certain threshold~\cite{meerson1,meerson2}.
A further confirmation that BBD-TC is essentially a ``bulk'' effect comes from
simulations~\cite{ramirez00} and theory~\cite{meerson1,meerson2} where
lateral walls are not required to observe it.  Convective circulation in
granular gases has been seen also in experiments, where lateral walls
are always inelastic~\cite{wildman01,lohse07,lohse10,windows13}, and
successive simulations with elastic~\cite{paolotti04} and also
inelastic walls~\cite{talbot02}.

The role of lateral walls in dilute granular convection has not been
fully understood yet. Some of the mentioned studies recognize that
lateral walls influence the observations. For instance, a downward
flow velocity is always observed near lateral walls, perhaps because
of a reduced buoyancy originated from enhanced
dissipation~\cite{ramirez00,wildman01,talbot02}. However, quite
contrasting results emerge in simulations and experiments concerning
the {\em necessity} of dissipation at the lateral walls. In some
simulations convection is strong even with elastic lateral
walls~\cite{ramirez00}, and in some experiments the threshold for
convection compares fairly well with theories where lateral walls are
absent~\cite{lohse10}. In other cases, convection is almost completely
killed when wall inelasticity goes to zero, a scenario - incompatible
with BBD-TC theories - seen both in simulations~\cite{talbot02} and in
experiments~\cite{windows13}.  Such discrepancies suggest that BBD-TC
is not the only mechanism able to generate convection in granular
gases.  Here we provide the evidence for TC in granular gases induced
by dissipative lateral walls (DLW). We employ
an ad-hoc experimental setup, able to isolate DLW-TC from BBD-TC,
together with molecular dynamics (MD) simulations.

Before introducing our experimental setup, we discuss a general
argument in favor of the existence of DLW-TC in \textit{any}
vertically vibrated granular system. Let us consider a 2D low density
gas (our analysis can be generalized to 3D systems) of identical solid
disks of mass $m$ enclosed by two inelastic parallel
walls. Perpendicular to the lateral walls, the bottom wall provides
energy to the system, for instance through steady vibration or (in
numerics) in the form of a thermostat. The system can be considered
closed by a fourth upper wall, or open: this does not change our
conclusion.  A constant gravity field $g$ is acting downwards along
the vertical ($y$) direction. 
For a dilute granular gas, $p=nT$ (with $p$ the pressure, $n$ the number
density, and $T$ the granular temperature)~\cite{D01}. An outgoing energy flux is always originated
at a dissipative wall~\cite{N99}, yielding in our case $\partial T /\partial x \neq 0$ at the lateral walls.
We wonder if such a gradient is compatible with hydrostatics, whose momentum balance reads
\begin{subequations}
\label{2d_balance}
\begin{align}
\partial_x p=\partial_x(n(x,y)T(x,y))=0\\
\partial_y p=\partial_y (n(x,y)T(x,y))=-mgn(x,y).
\end{align}
\end{subequations}
According to the first equation $p(x,y) \equiv p(y)$, which, used in the second equation,
forces $n(x,y) \equiv n(y)$ and also $T(x,y)\equiv
T(y)$. However, this contradicts the necessity of a horizontal
temperature gradient induced by DLW. 
Thus, in the system described any steady state must have flow (that is, a macroscopic velocity field $u \neq 0$) and, 
since the system is closed in the horizontal, the flow will be convective.

\begin{figure}[tb!]
\includegraphics[width=8cm]{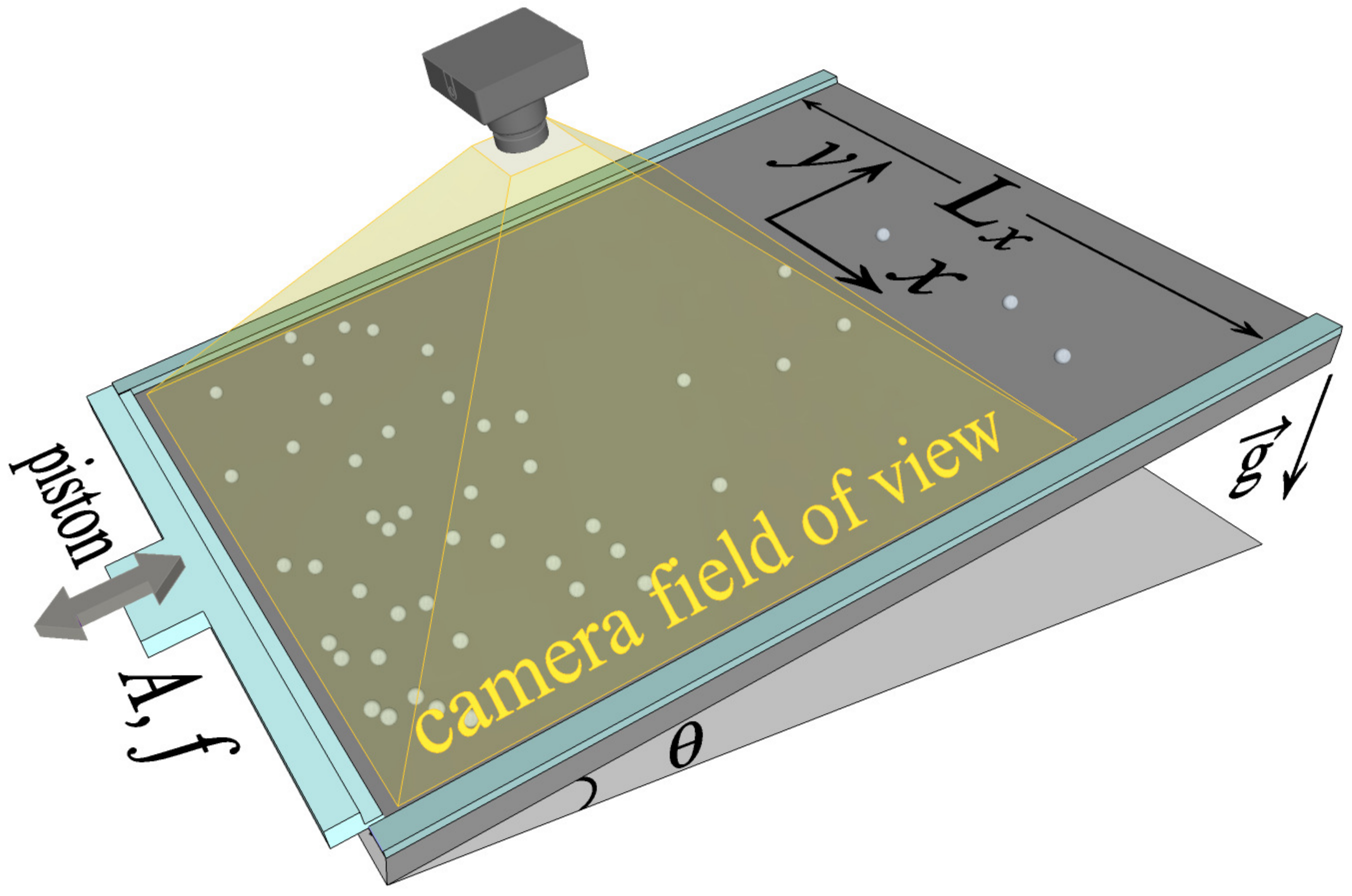}
\caption{Sketch of the experimental setup. The real length (along $y$) of the inclined plane, $L_y \gg L_x$, is not marked as it is not essential in the description of the system. \label{fig:setup}}
\end{figure}

Inspired by such a straightforward observation, we have set up a
granular gas experiment under low-gravity, sketched in
Fig.~\ref{fig:setup} and detailed in Supplemental
Material~\cite{sm}. In theoretical studies~\cite{meerson1,meerson2} it
is seen that when $g \to 0$ the BBD-TC instability requires a larger
and larger wavelength to develop: therefore at a given width of the
system there is a gravity value under which unstable perturbations
cannot appear and the BBD-TC is suppressed.  Our setup is a gas of $N$
spherical steel beads (diameter $d=1~\mathrm{mm}$) moving inside a
cuboid of dimensions $L_x=175~\mathrm{mm}$, $L_y=600~\mathrm{mm}$ and
$L_z=1.5~\mathrm{mm} = 1.5 d$, thus assuring for a quasi-2D dynamics
restricted to the $xy$ plane. This plane has a tilt angle $\theta$
with respect to the horizontal: the spheres therefore move with an
effective value of gravity $g_{eff} \approx (5/7) g \sin(\theta)$,
where $g$ is Earth's gravity acceleration and the constant $5/7$ is
due to the moment of inertia of spheres (at the chosen values of
$\theta$ our trajectories are dominated by pure rolling, see
also~\cite{blair03}).  The limits of the plane consist of two lateral
``walls'' made of polycarbonate (at $x=\pm L_x/2$), one inferior wall
(at $y=0$) consisting of a vibrating
Plexiglas\textsuperscript{\textregistered} piston, and a far top side
which is also made of polycarbonate at $y=L_y$. The bottom plate (area
$L_x\times L_y$) is made of aluminum alloy and finally the system is
covered with a glass plate. The piston oscillates with amplitude $A$
and frequency $f$. The average squared velocity of the piston, here
defined as $v_0^2=(A 2\pi f)^2/2$, helps setting the energy and
velocity units in the following.  Our setup is similar to that of
previous experiments~\cite{kudrolli97b,kudrolli00} and is of the kind
described by eqs. \eqref{2d_balance}. It is worth noting that here,
contrary to more common vertical
setups~\cite{wildman01,lohse07,lohse10,windows13}, we can reduce and
control the effective gravity.  A high speed camera records square
images of size $L_x \times L_x$ starting from the maximum position of
the piston, i.e. excluding the topmost very dilute region. A pairwise
acquisition protocol (see~\cite{sm}) allows us to reconstruct the
average fields $u(x,t)$, $n(x,t)$ and $T(x,t)$ (flow velocity,
particle density, and granular temperature, as usually
defined~\cite{poeschel}) in the visible field, with a a $40 \times 40$
mesh. A preliminary study of non-interacting trajectories has
confirmed the value of $g_{eff}$ and has shown the presence of small
frictional effects in the form of both Coulomb-like friction and
viscous-like frictions~\cite{sm}.
\begin{figure*}
\includegraphics[width=4.4cm,height=4.4cm,angle=-90]{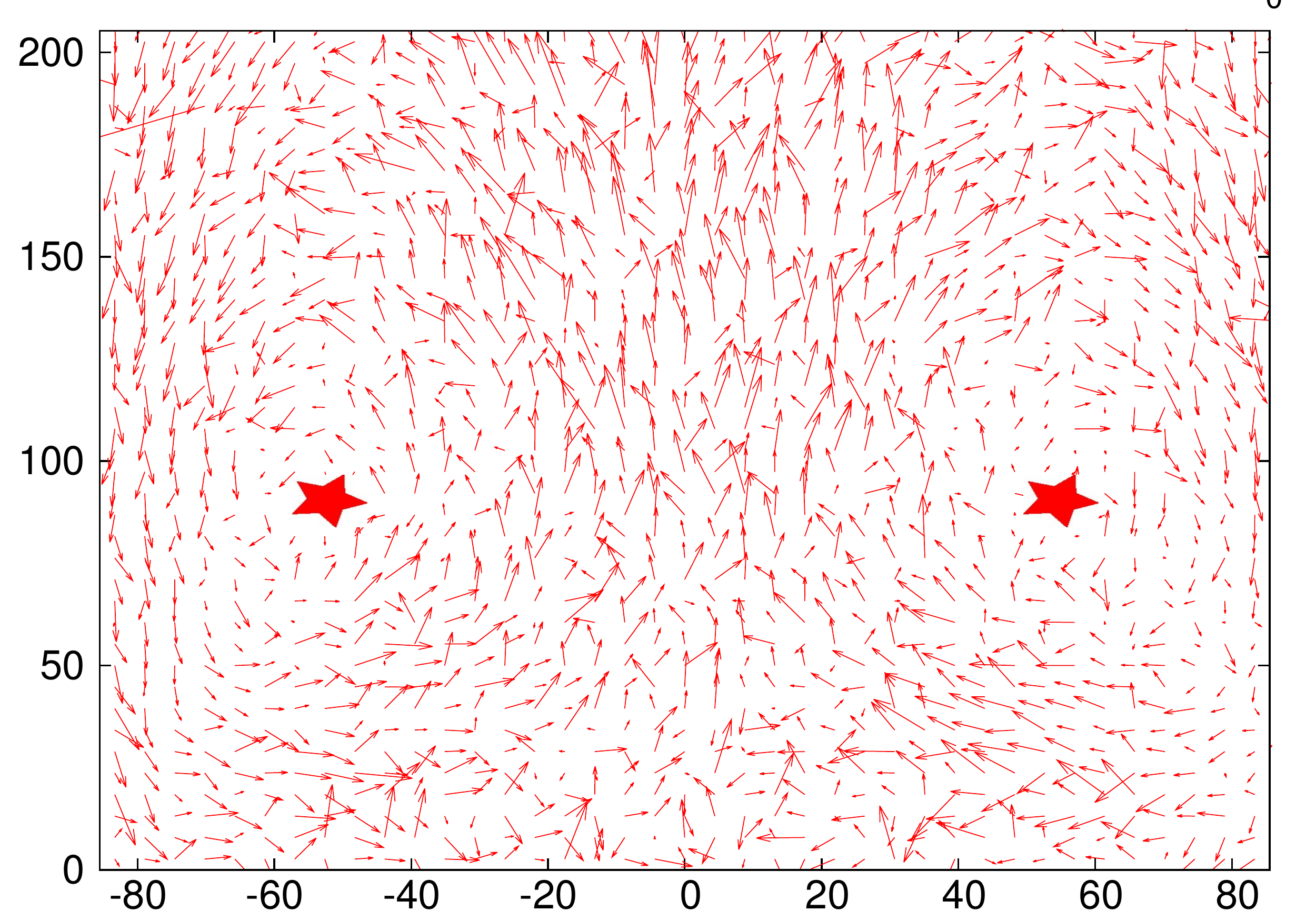}
\includegraphics[width=4.4cm,height=4.4cm,angle=-90]{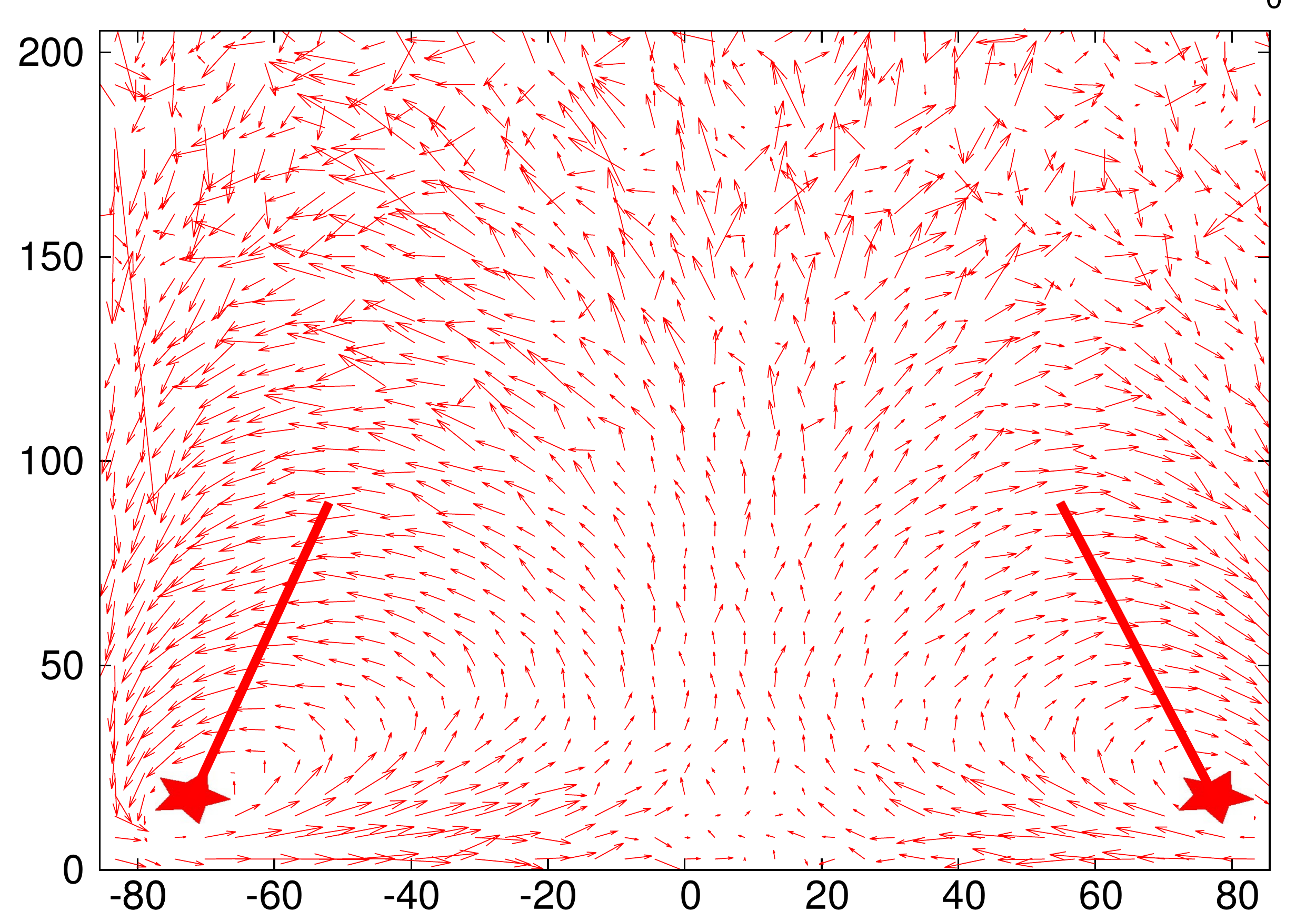}
\includegraphics[width=4.4cm,height=4.4cm,angle=-90]{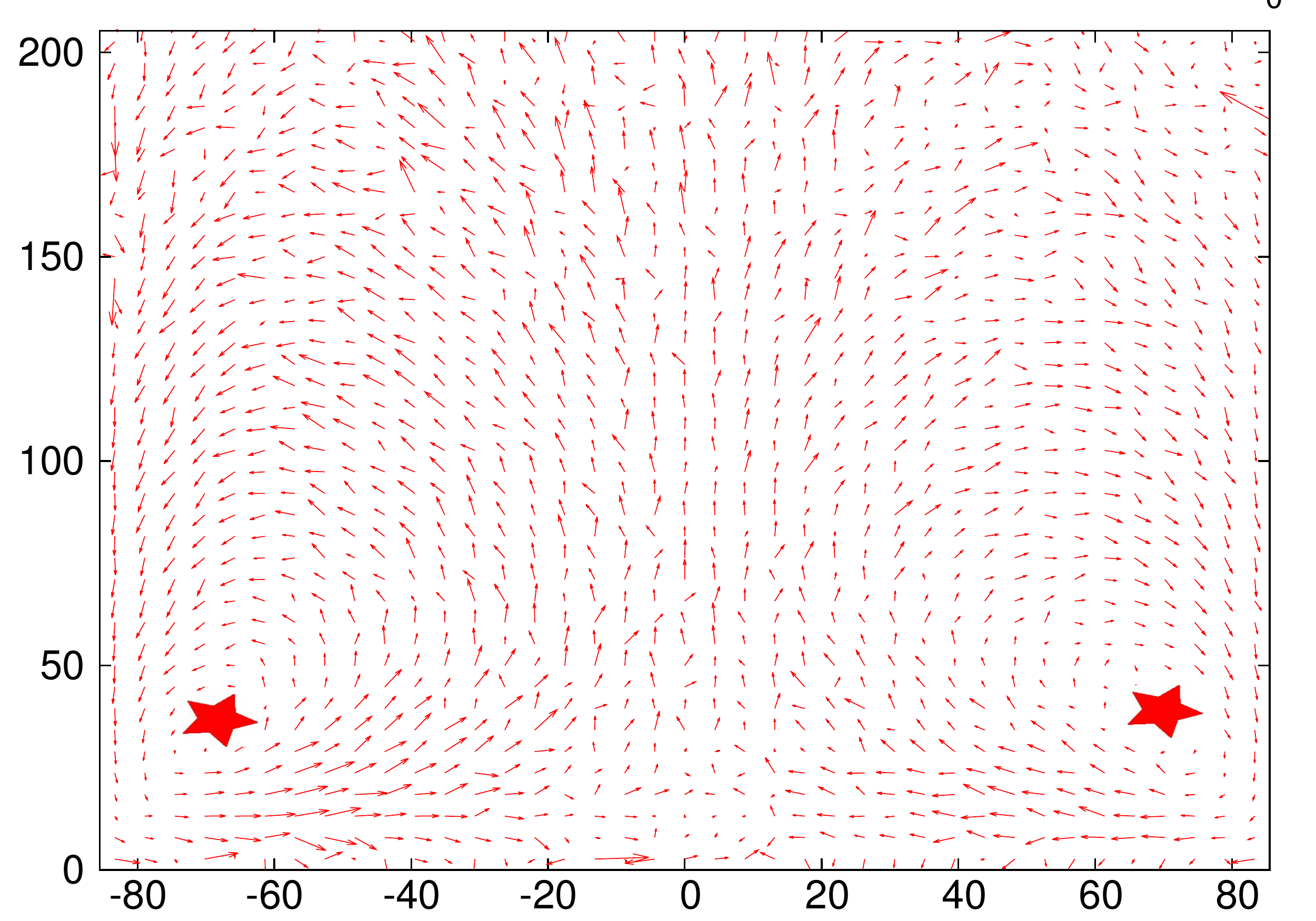}
\includegraphics[width=4.4cm,height=4.4cm,angle=-90]{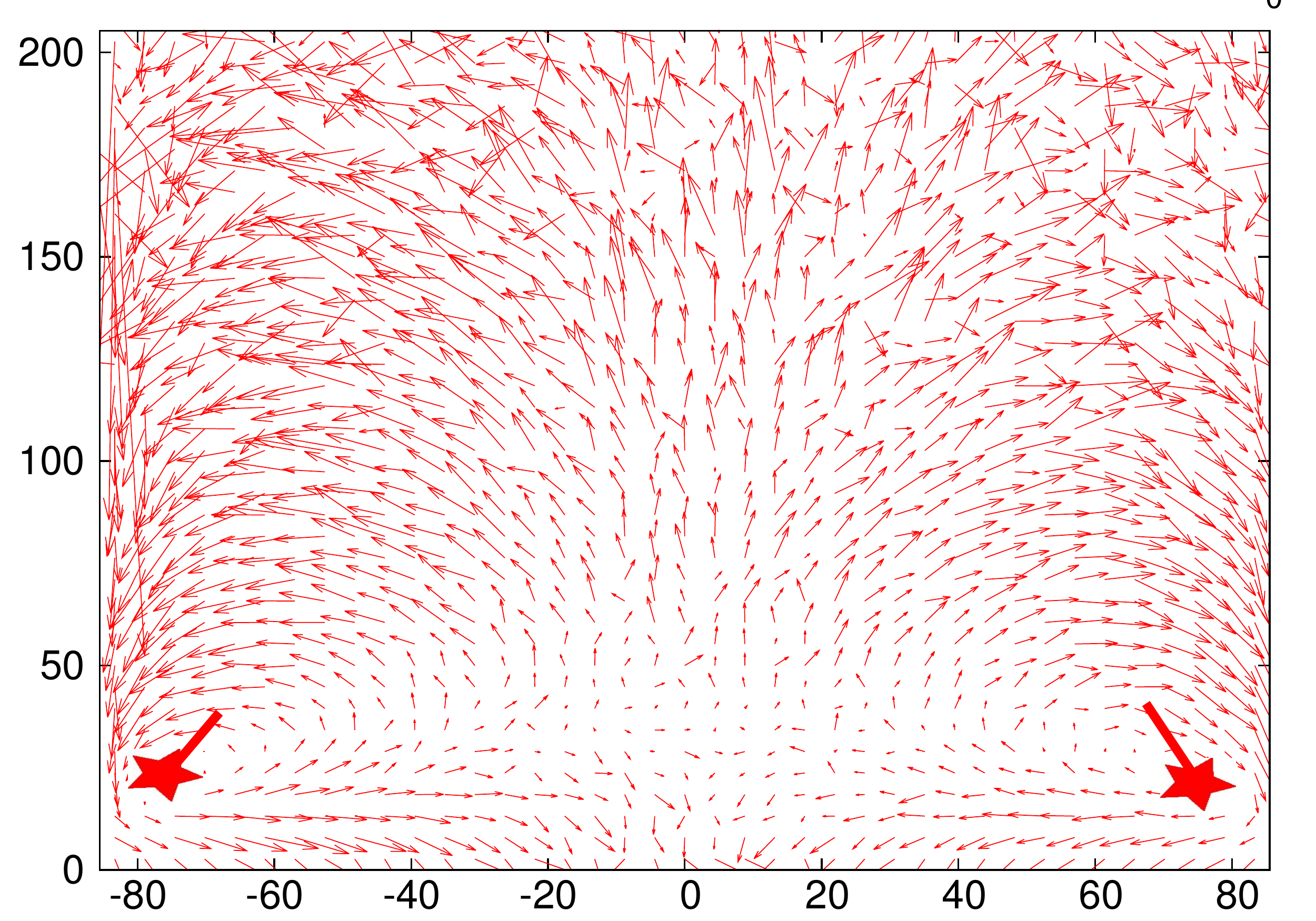}\\
\includegraphics[width=2.2cm,height=2.15cm,angle=-90]{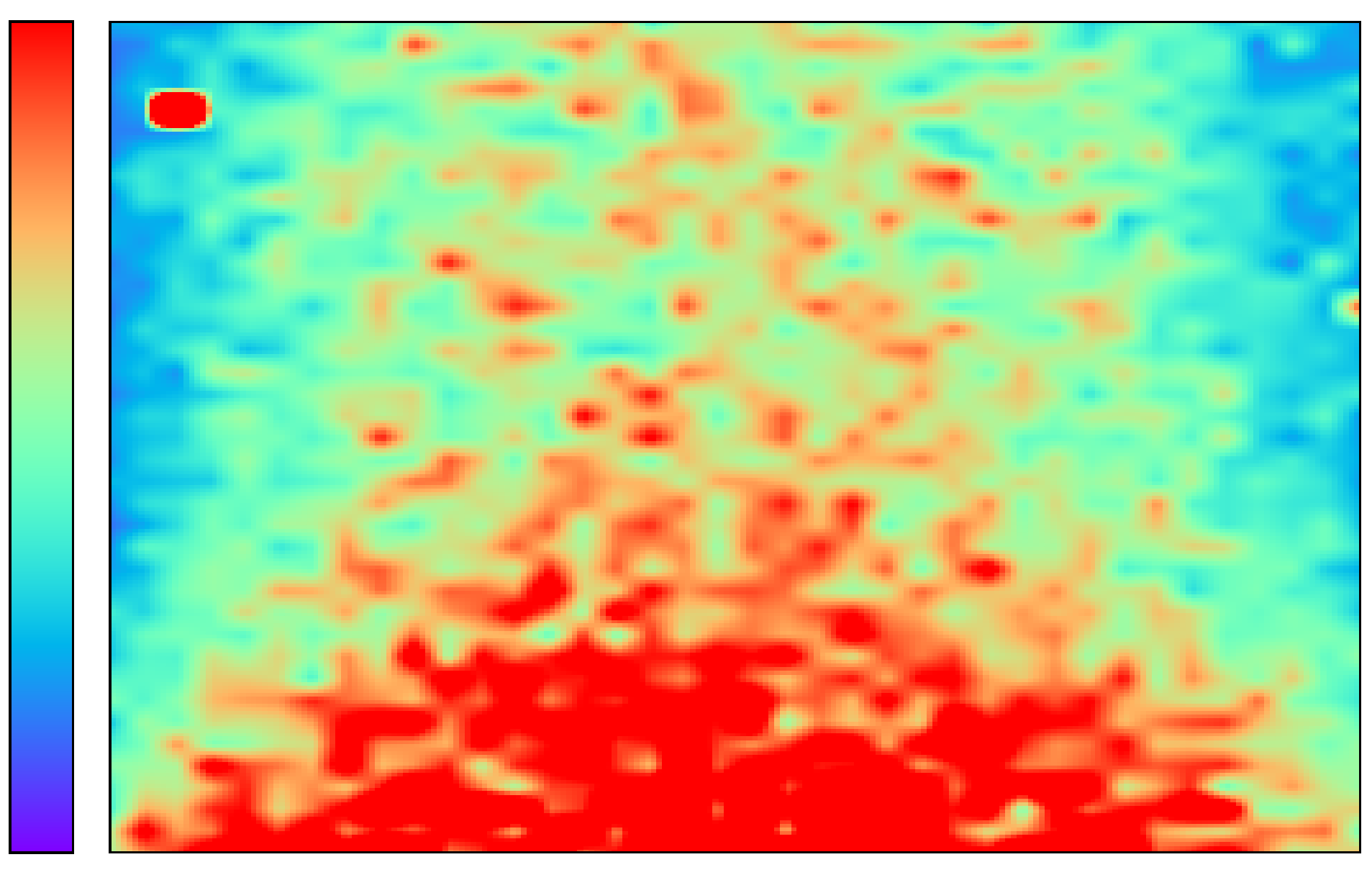}
\includegraphics[width=2.2cm,height=2.15cm,angle=-90]{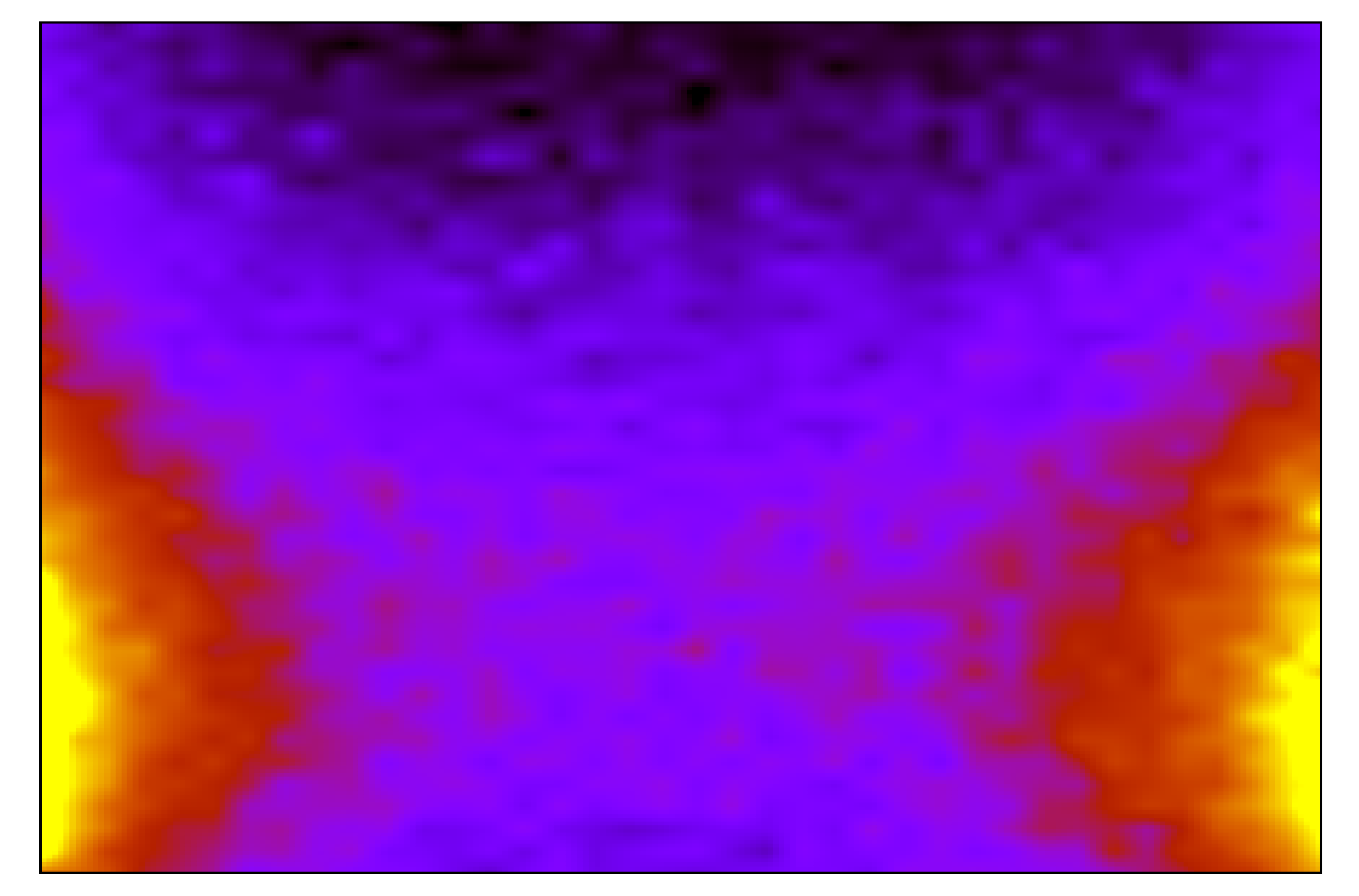}
\includegraphics[width=2.2cm,height=2.15cm,angle=-90]{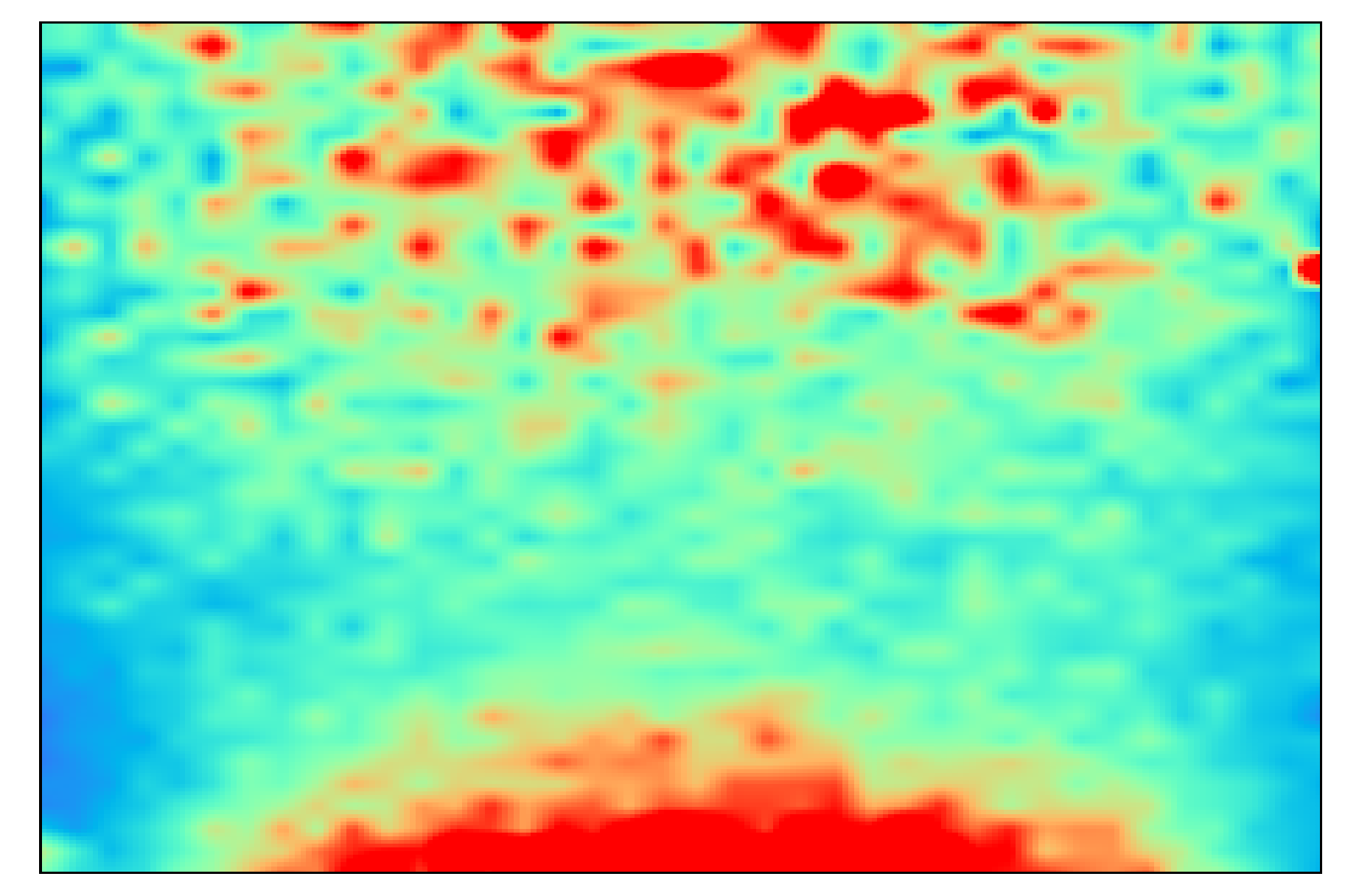}
\includegraphics[width=2.2cm,height=2.15cm,angle=-90]{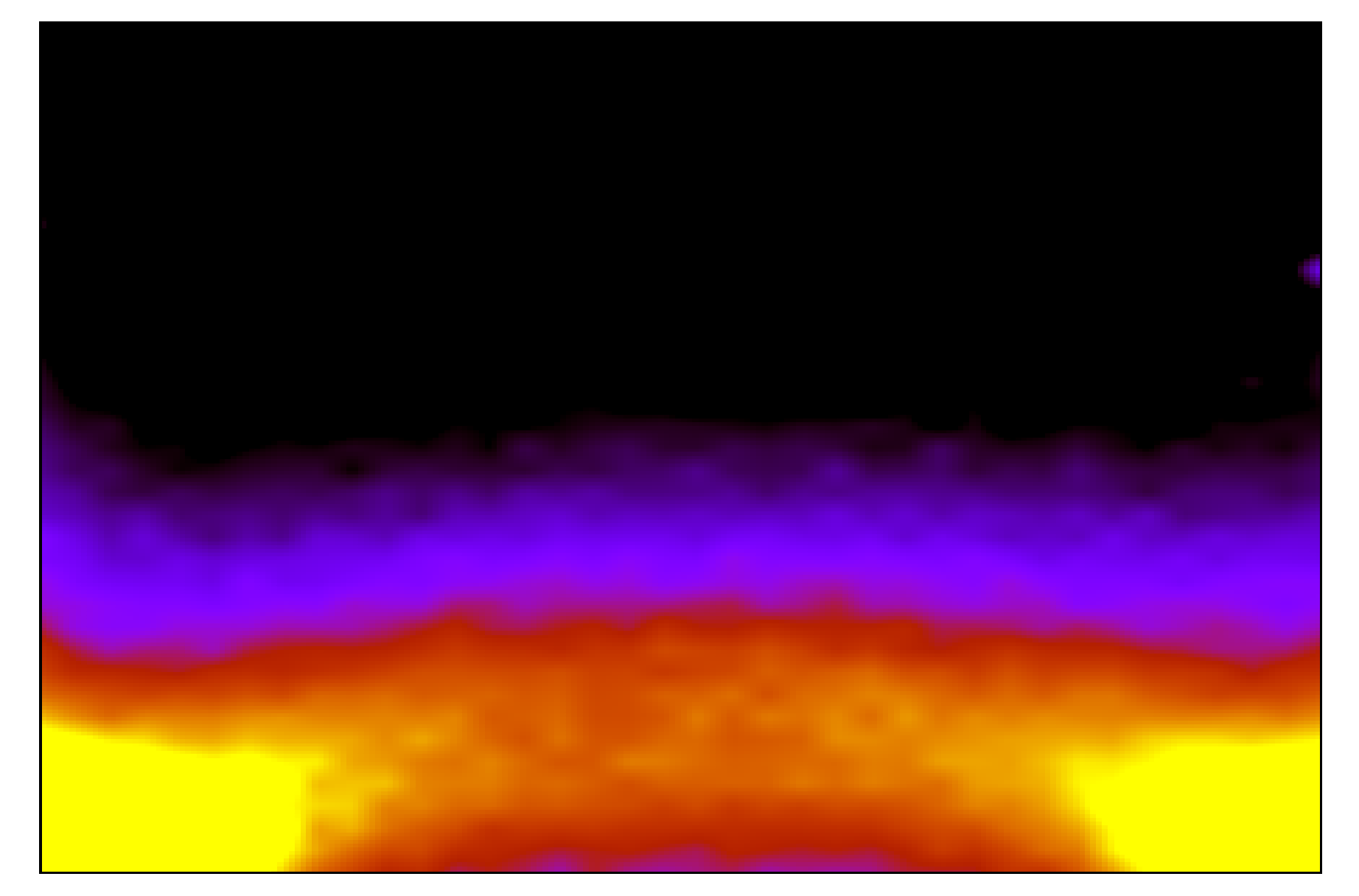}
\includegraphics[width=2.2cm,height=2.15cm,angle=-90]{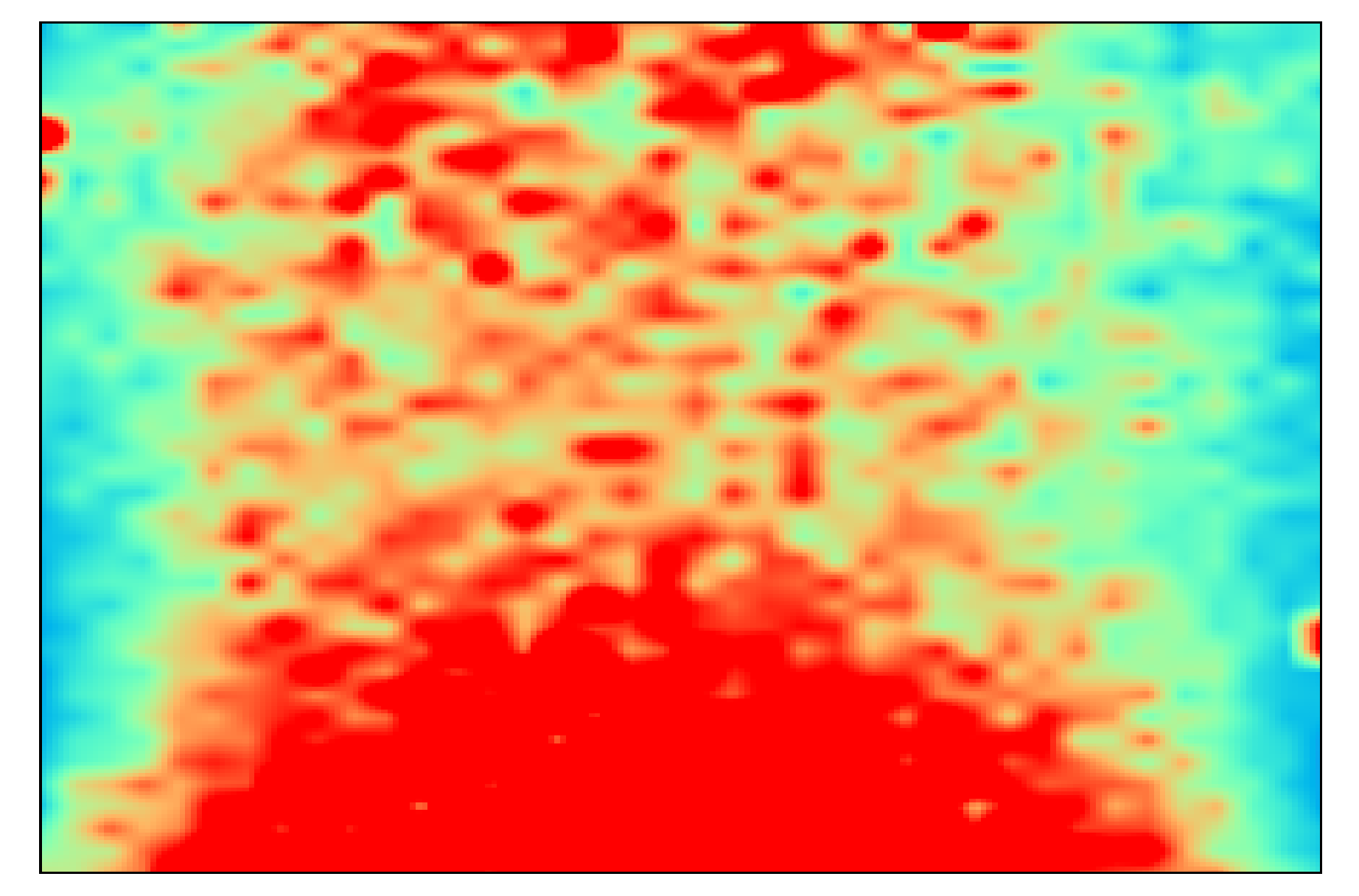}
\includegraphics[width=2.2cm,height=2.15cm,angle=-90]{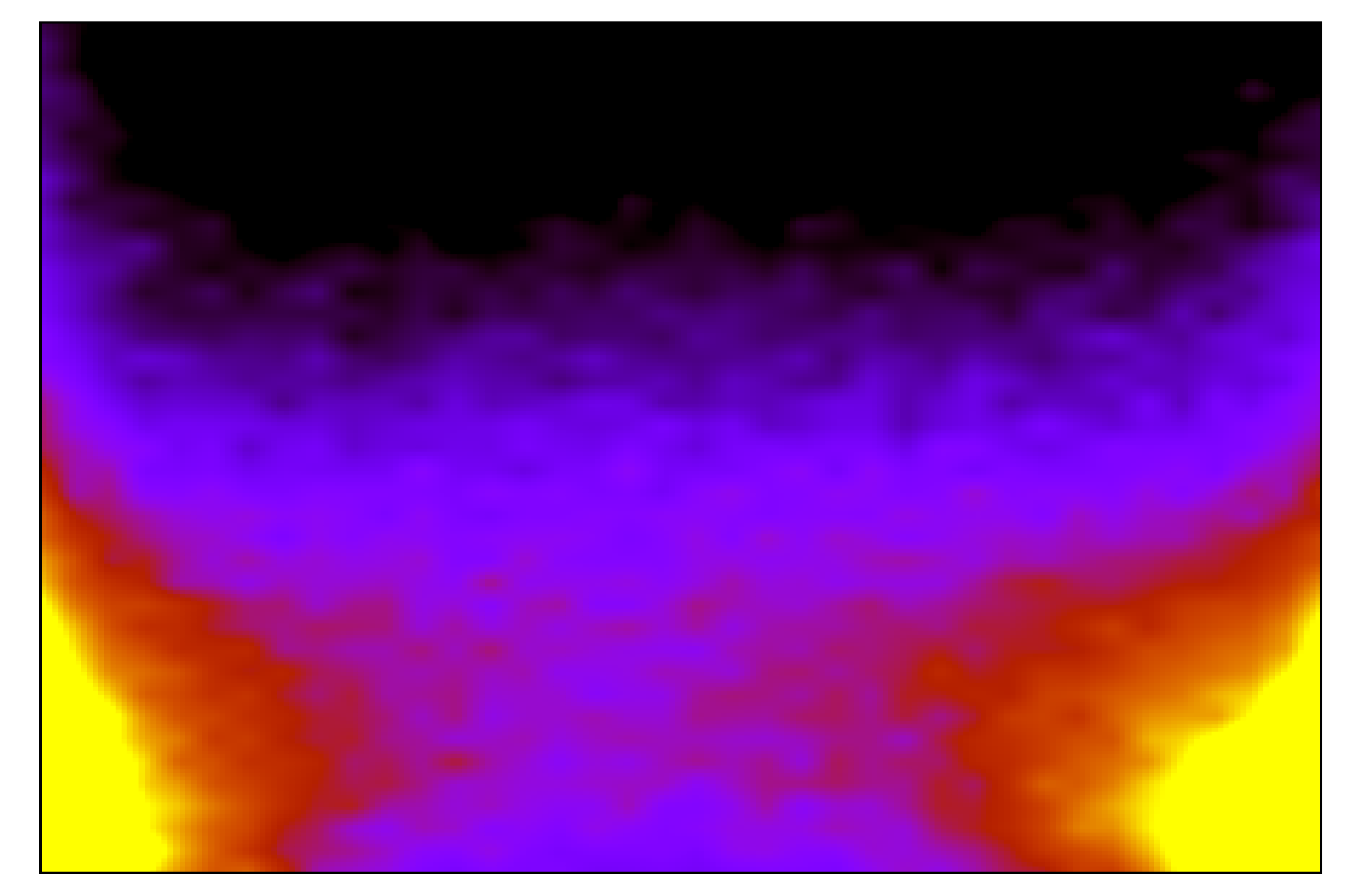}
\includegraphics[width=2.2cm,height=2.15cm,angle=-90]{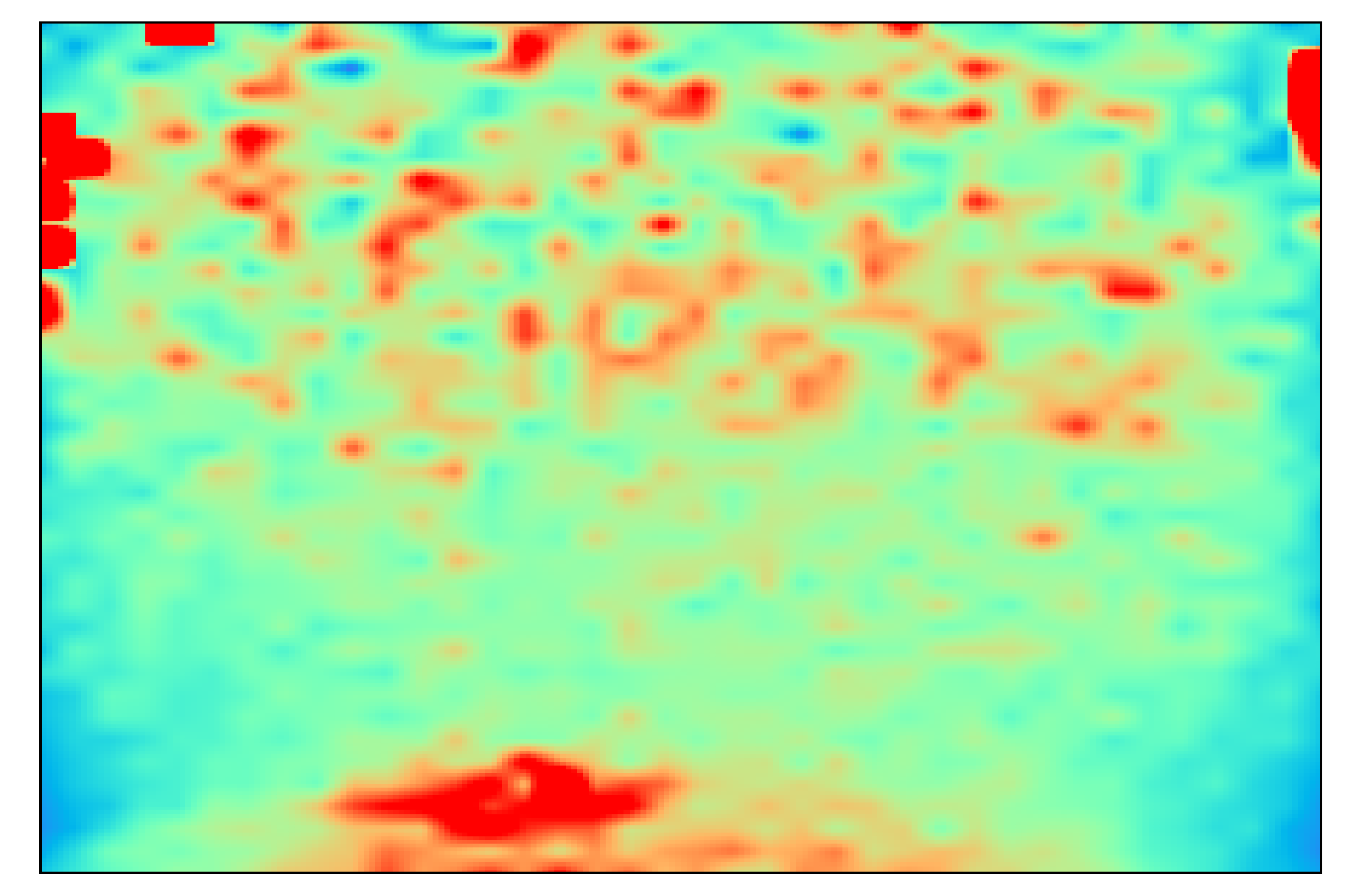}
\includegraphics[width=2.2cm,height=2.15cm,angle=-90]{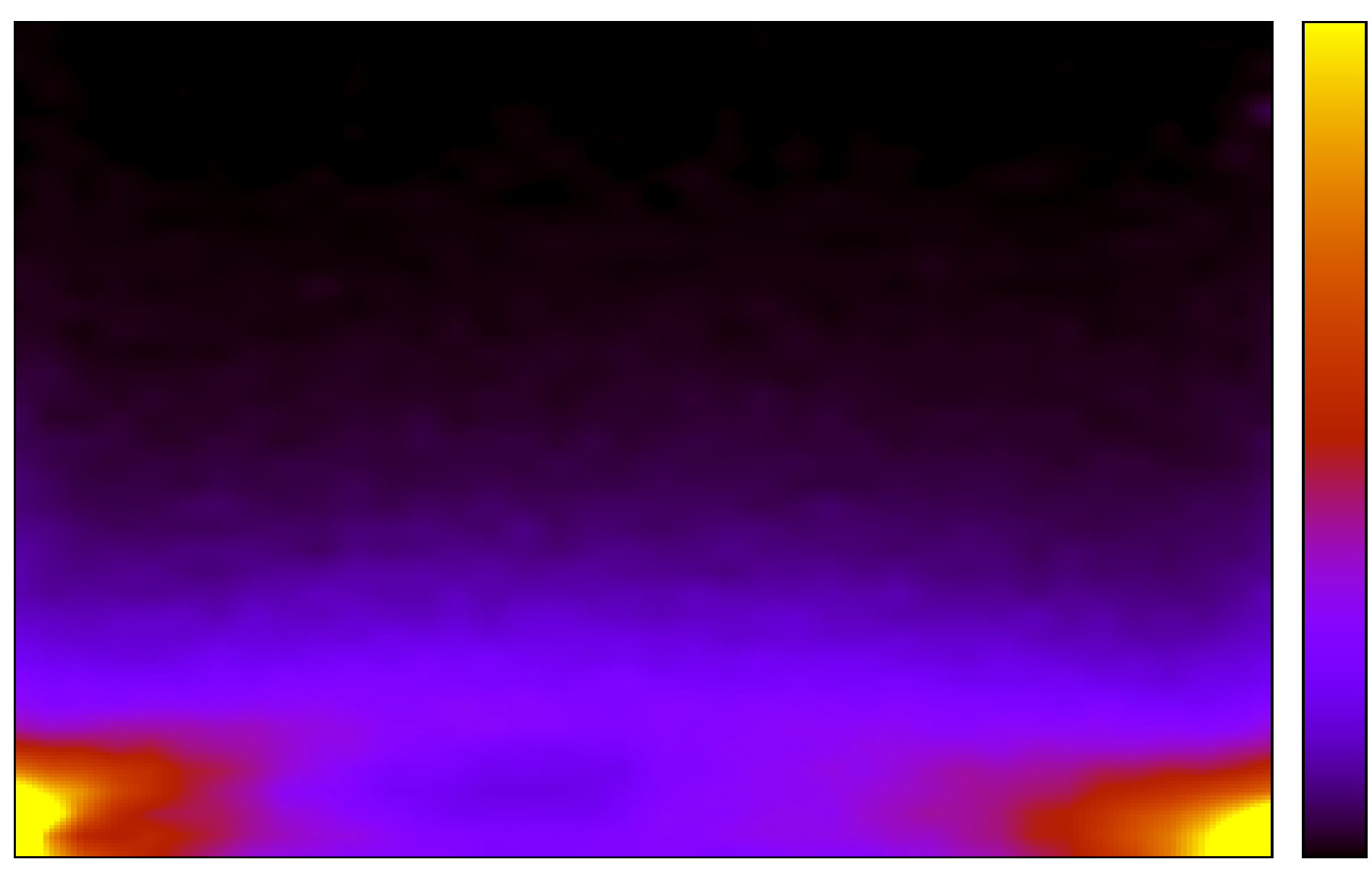}
\caption{ Fields from four experiments with $f=45$ Hz, $A=1.85$ mm
  ($v_0=370$ mm/s) and different values of $N$ and $g_{eff}$. Top row: the
  velocity field with the two convective cells; coordinates are given in units of particle's diameters ($=1$ mm); the center of each
  cell is marked by a star; in graphs B and D a thick line represents
  the movement of the cell's center with respect to case A and C
  respectively. Bottom row: each couple of graphs show the (mass-free) temperature
  $T/m$ (left) and local packing fraction $\nu=n \pi (d/2)^2$ (right)
  fields corresponding to the case in the top row. The $T$ scale (see legend on the left) goes
  from black (colder) at $T=0$ to red (hotter) at $T=T_{max}$. The
  $\nu$ scale goes from black (more dilute) at $\nu=\nu_{min}$ to yellow
  (denser) at $\nu=\nu_{max}$. Values of $T_{max}/mv_0^2$ are: A) $0.2$, B) $0.1$, C) $0.2$, D) $0.4$.
Ranges for ($\nu_{min},\nu_{max}$) are: A) $(0.02 \%, 0.15 \%)$, B) $(0.02 \%, 0.15\%)$, C) $(0.04 \%,0.1 \%)$, D) $(0.005 \%,0.5 \%)$.}
\label{fig:varioN_2d}
\end{figure*}
We used $N \in [100,1500]$,
which yields an average 2D packing fraction $\nu_{2D}=N\pi (d/2)^2/(L_xL_y)$
in the range of $\nu_{2D} \in [0.1 \%, 1 \%]$, with observed local
variations reaching up to $\sim 5\%$ at the highest values of $N$ and
$g_{eff}$: in summary, we are always in the dilute regime. A detailed assessment of mean free paths measured in the system is
discussed in~\cite{sm}. The amplitude of vibration has been fixed to
$A=1.85$ mm, while the frequency is varied in the range $f \in [10, 45]$
Hz. The inclination angle is varied in the range $\theta \in [0.011,
  0.130]$ radiants. Explored values of the rescaled maximum
acceleration are $\Gamma=A(2\pi f)^2/g_{eff} \in [70,3800]$.  

Let us discuss our experimental results. In all performed experiments,
with an exception discussed below, we always observed convection with
two convective cells that span the full width of the 2D
plane. Examples of the experimental 2D velocity fields are shown in
the top row of Fig.~\ref{fig:varioN_2d}.  In the special case $N=100$
we have not observed convection. This may be due to fact that the mean
free path is larger than $L_x$ (Knudsen gas) and a consequent possible
breakdown of the ideal gas equation of state.  In the lower graphs of
Fig.~\ref{fig:varioN_2d} we show the experimental hydrodynamic
$2\mathrm{D}$ fields for density and temperature.  The fields clearly
display gradients in both $x$ and $y$ directions.  The temperature
field (blue-red graphs in Fig.~\ref{fig:varioN_2d}) tends to decrease
when moving along $x$ from the center to the lateral boundaries as
expected from the simple argument of an outward energy flow due to
DLW. Along $y$, the granular temperature shows a richer behavior: for
low $N$ and $g_{eff}$ the temperature is dominated by a negative
gradient. This would be associated with heat transport from the bottom
thermostat to the upmost cold region, where energy is continuously
dissipated by inelastic particle-particle and DLW-particle
collisions. At larger $N$ and $g_{eff}$ a temperature minimum is
observed along $y$, which can be explained by granular hydrodynamics
taking into account a {\em secondary} energy flux which is associated
with the density gradient~\cite{soto99,brey01d}. The density field
(black-yellow graphs in Fig.~\ref{fig:varioN_2d}) displays a
saddle-like structure. On a horizontal line it shows its largest
values near the two DLW ($x=\pm L_x/2$) and a minimum halfway, i.e. at
$x=0$. This is consistent with the fact that particle-wall inelastic
collisions favor condensation near the walls.  Along the $y$
direction, on the contrary, the density shows a maximum at some given
height. Such a maximum shifts toward the base as $N$ or $g_{eff}$
increase, a fact that is consistent with the increase of steepness in
the decay of temperature in the lowest region of the gas. A key
observation concerns the behavior of the center of the convective
cells: they appear to move toward the lowest corners when $N$ or
$g_{eff}$ are increased, as highlighted by the stars superimposed to
the velocity fields in Fig.~\ref{fig:varioN_2d}B
and~\ref{fig:varioN_2d}D. Variations
occurring when $N$ is increased (decreased) are qualitatively similar
to those occurring when $g_{eff}$ is increased (decreased).

\begin{figure}[b!]
\includegraphics[height=4.25cm,width=4.4cm,clip=true,angle=-90]{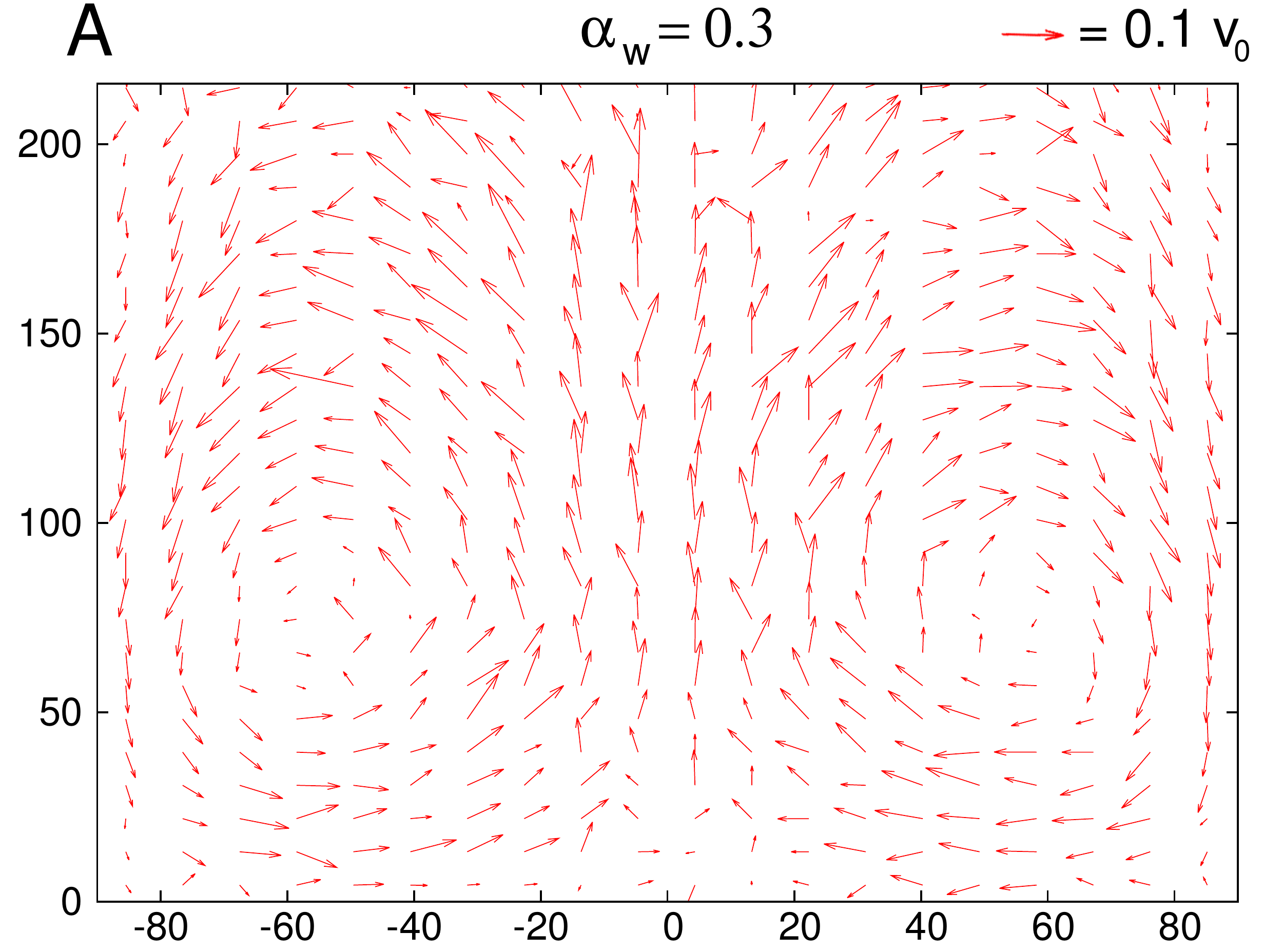}
\includegraphics[height=4.25cm,width=4.4cm,clip=true,angle=-90]{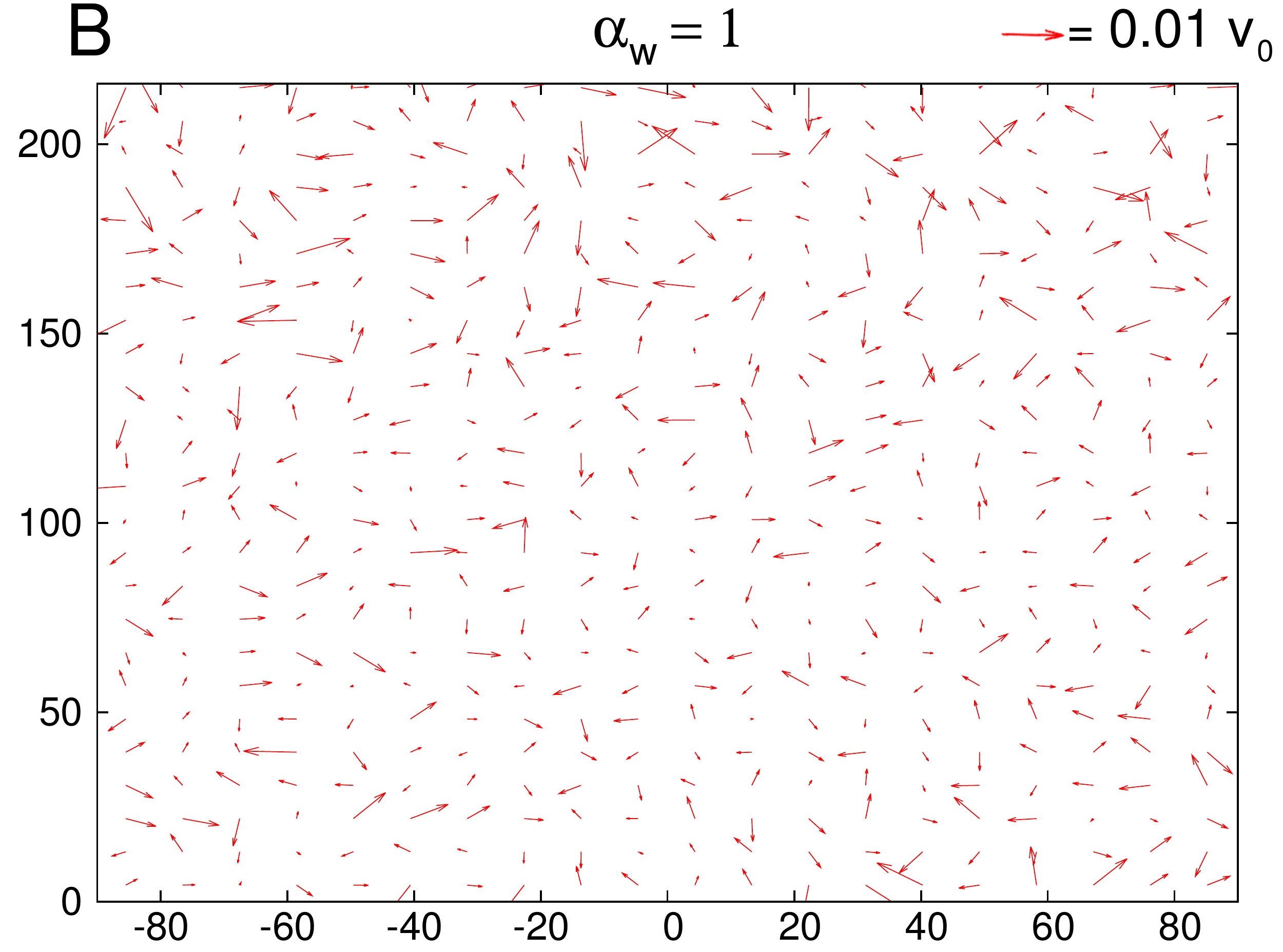}\\
\includegraphics[height=2.05cm,width=2.1cm,clip=true,angle=-90]{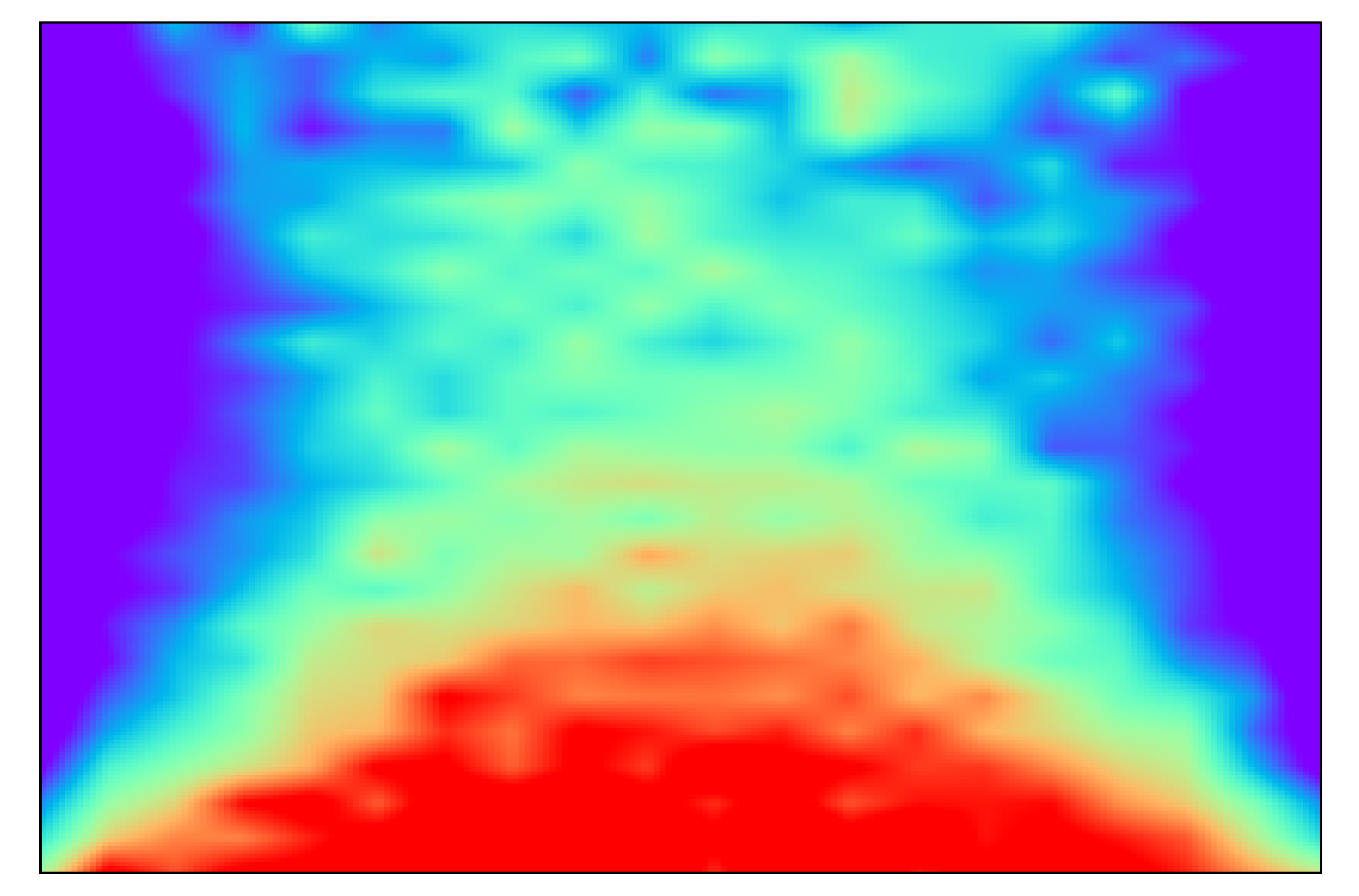}
\includegraphics[height=2.05cm,width=2.1cm,clip=true,angle=-90]{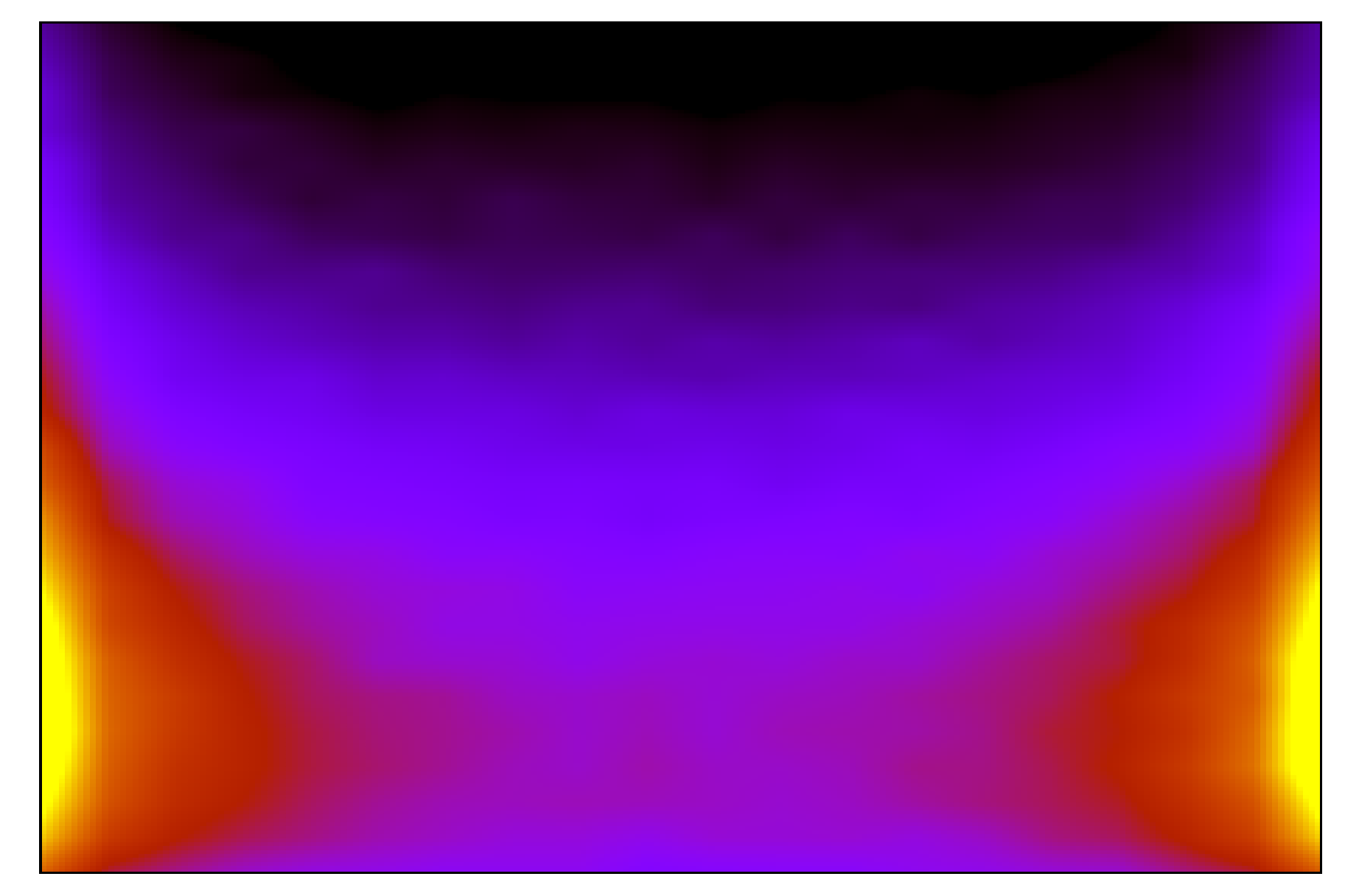}
\includegraphics[height=2.05cm,width=2.1cm,clip=true,angle=-90]{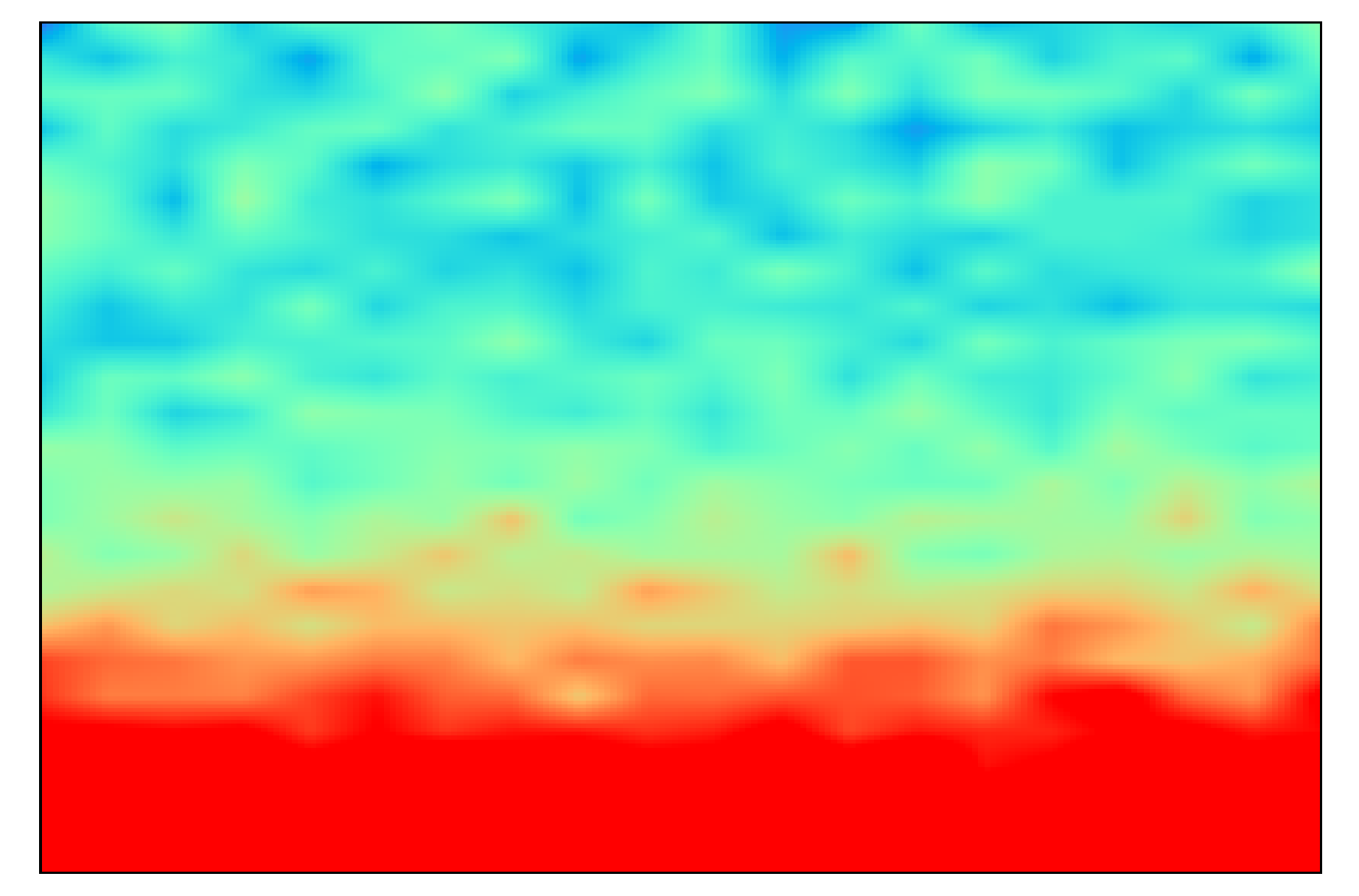}
\includegraphics[height=2.05cm,width=2.1cm,clip=true,angle=-90]{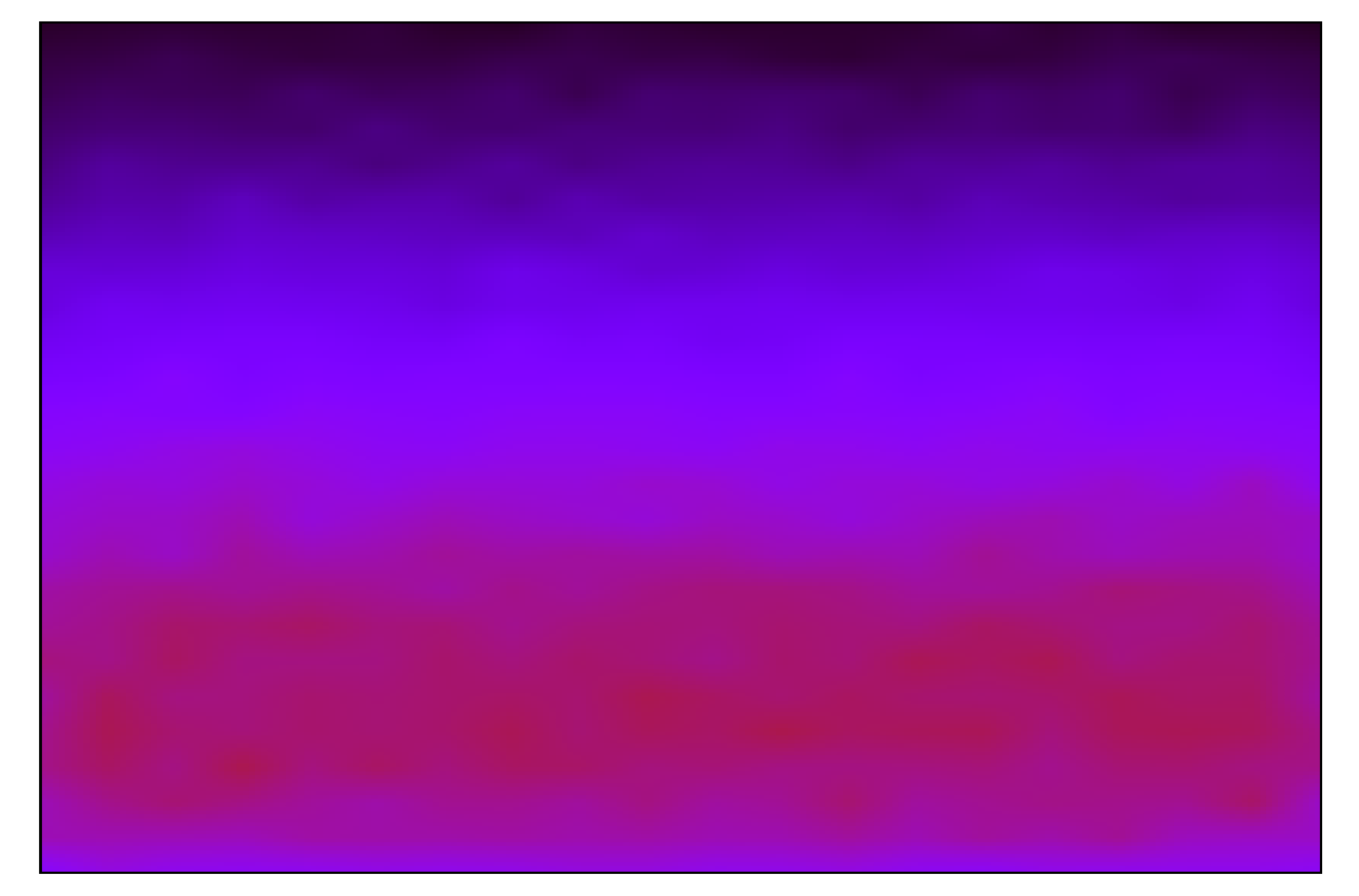}
\caption{Fields from two simulations with $N=500$, $g_{eff}=0.016 g$,
  $\alpha=0.98$, $v_0=280$ mm/s and different values of $\alpha_w$: in
  A $\alpha_w=0.3$, in B $\alpha_w=1$ (elastic walls).  Coordinates and colors are the same as in Fig.~\ref{fig:varioN_2d} with
the following values of $T_{max}/mv_0^2$: A) $0.1$, B) $0.2$; and the following ranges for ($\nu_{min},\nu_{max}$): A)
  $(0.05 \%, 0.5 \%)$, B) $(0.05 \%, 0.5\%)$.\label{fig:sim}}
\end{figure}

\begin{figure}
\includegraphics[width=3.cm,height=8.5cm,clip=true,angle=-90]{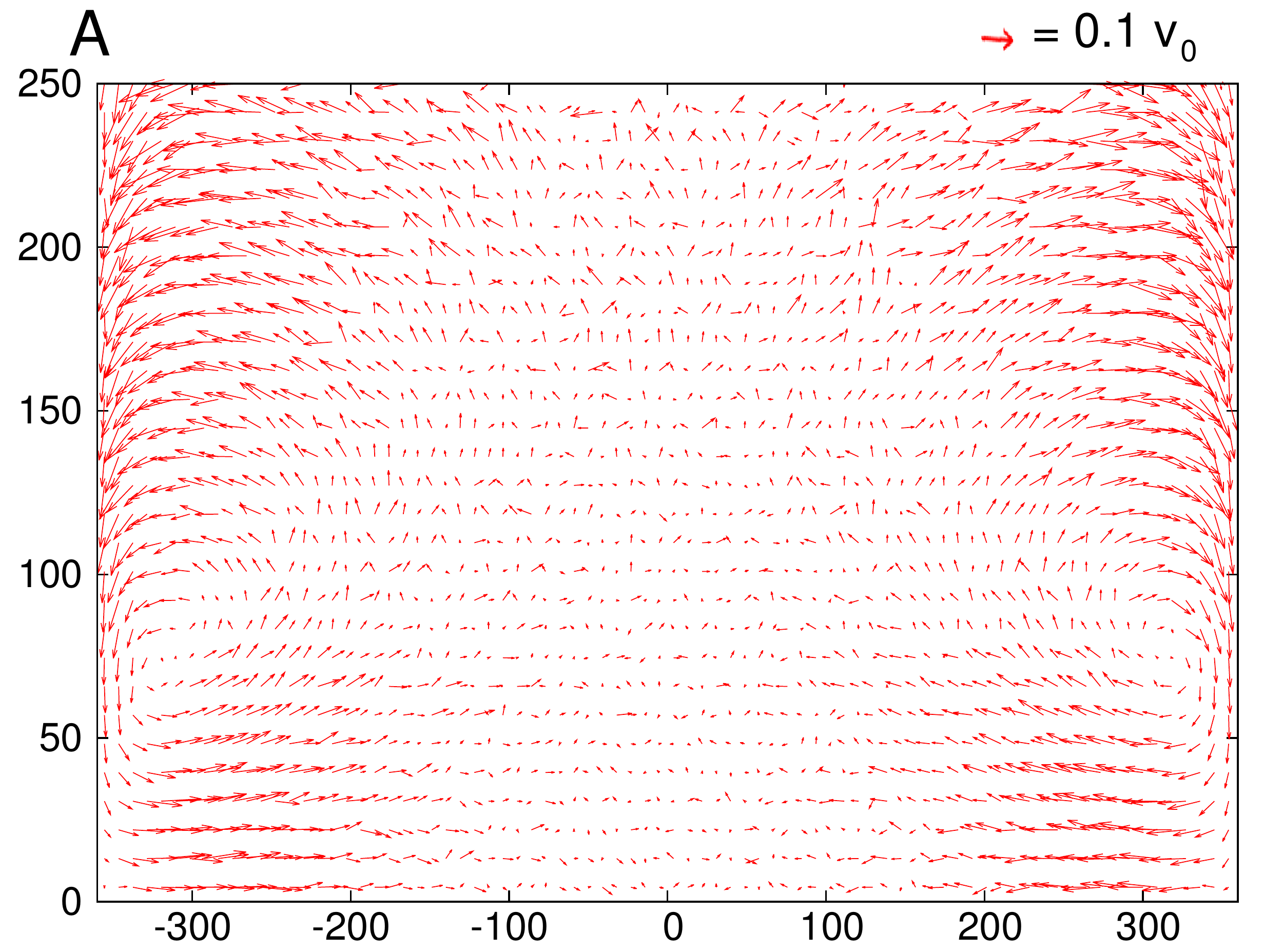}\\
\includegraphics[width=4cm,height=3cm,clip=true]{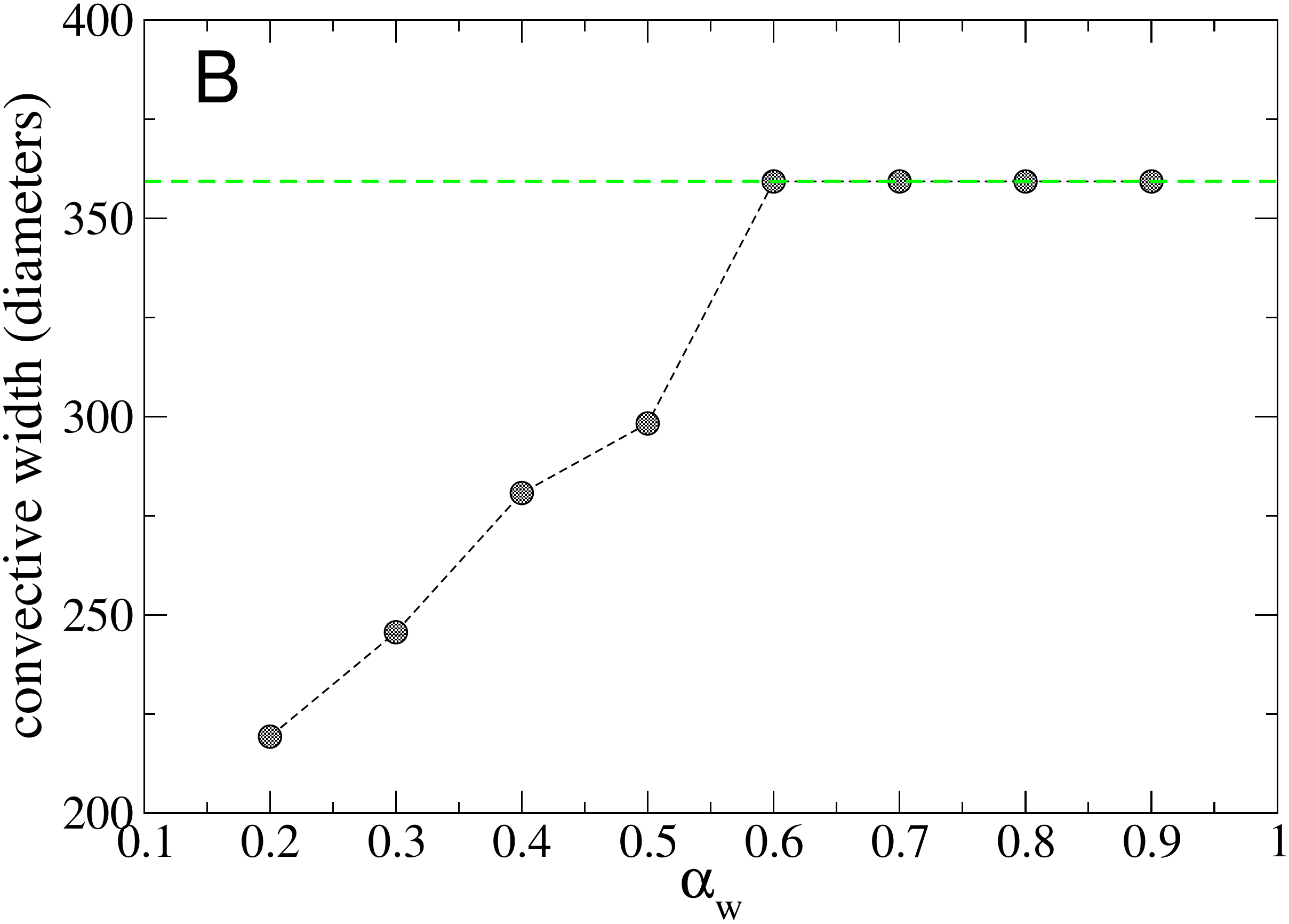}
\includegraphics[width=4cm,height=3cm,clip=true]{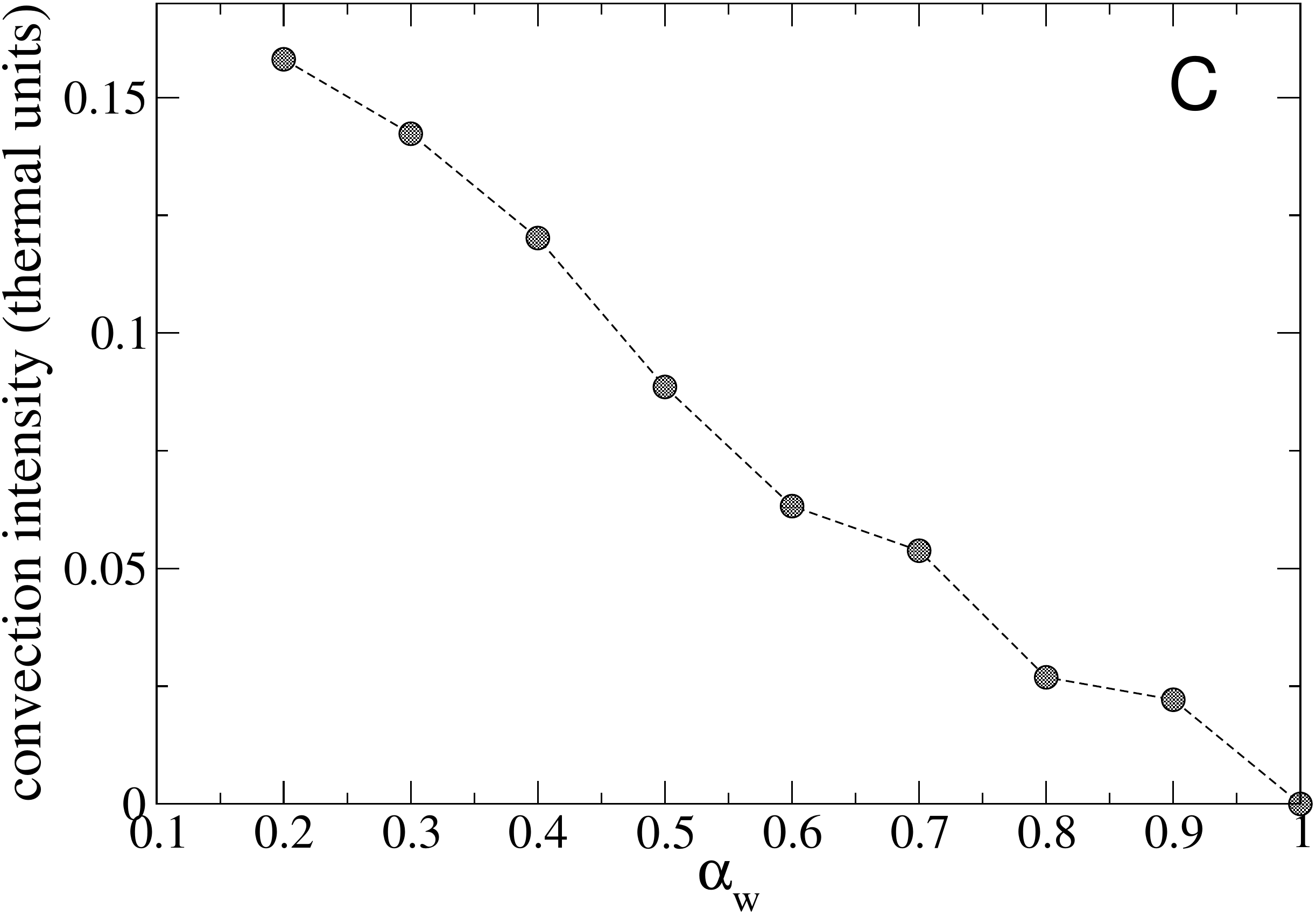}\\
\includegraphics[width=4cm,height=3cm,clip=true]{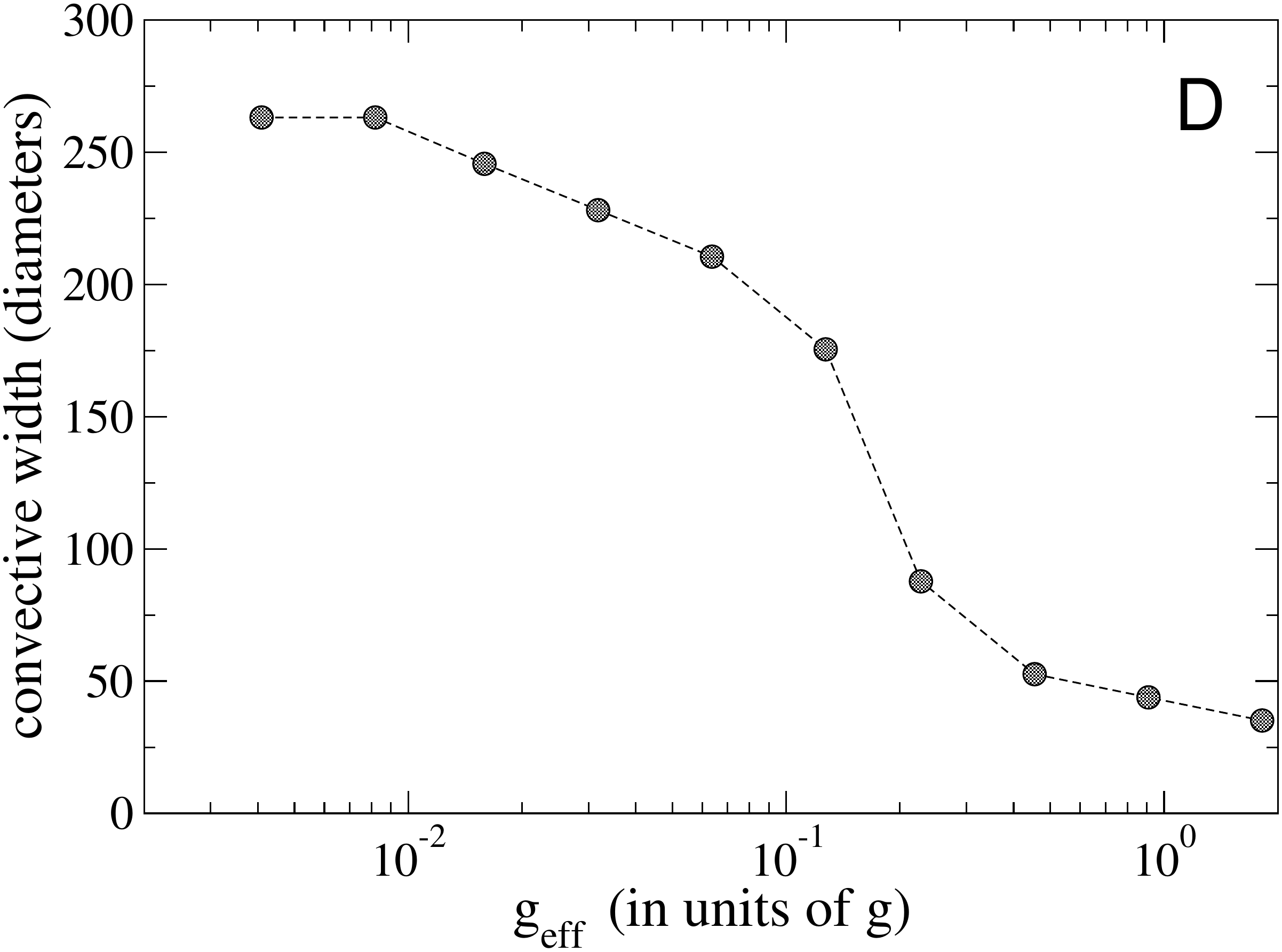}
\includegraphics[width=4cm,height=3cm,clip=true]{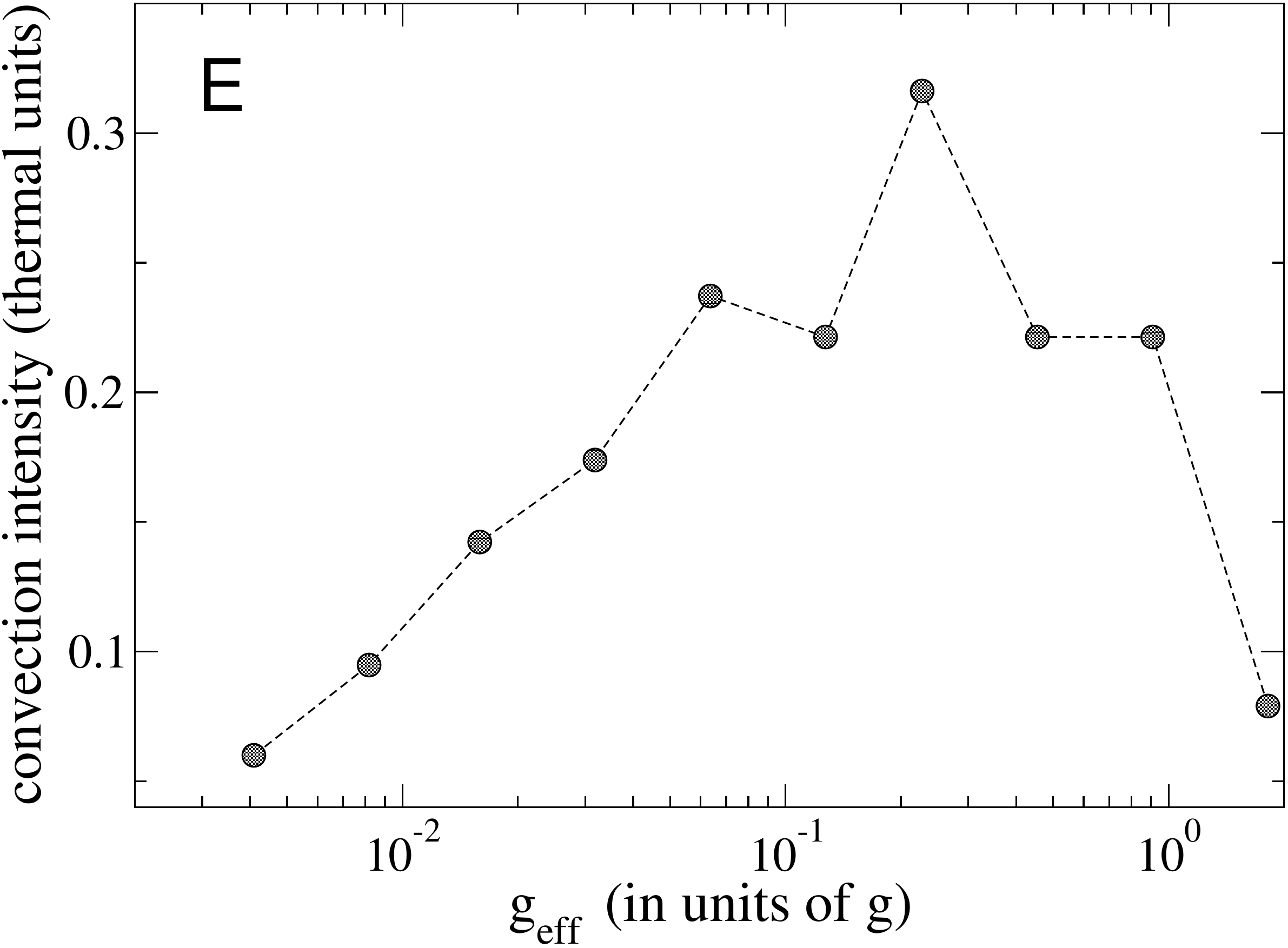}
\caption{A: velocity field of a large system ($L_x=720$ diameters)
  with $N=2000$, $v_0=280$ mm/s, $\alpha=0.98$, $\alpha_w=0.3$, and
  $g_{eff}=0.016 g$. B and C: width and intensity, respectively, of
  convection as a function of $\alpha_w$ with $g_{eff}=0.016g$. D and E: width and intensity, respectively, of
  convection as a function of $g_{eff}$ with $\alpha_w=0.3$. In graphs B-E we used $L_x=720$,
  $N=2000$, $v_0=280$ mm/s, $\alpha=0.98$. \label{fig:width}}
\end{figure}

How do our observations compare with previous studies? It is likely
that experiments in~\cite{windows13} and simulations
in~\cite{talbot02} were dominated by DLW-TC, as
comparison of those results with our Figures~\ref{fig:sim} and~\ref{fig:width} (below)
demonstrate. Both cases concern 3D systems, but our argument after
Eq.~\eqref{2d_balance} is not affected by a third dimension.
In~\cite{lohse07} it has been shown that in a dilute granular fluid
under vertical vibration, convection takes place only in a limited
region of parameter space, at variance with our general argument
showing that DLW-TC should always be observed. Our conjecture is that
for larger and larger $g_{eff}$ (or $N$) the two DLW-TC cells occupy a
smaller and smaller region of the system, up to a point where the
DLW-TC cell is so tiny that it could go unnoticed. As a matter of
fact, all experiments in~\cite{lohse07} are performed at Earth's
gravity, which is much larger than our $g_{eff}$.

Since our setup has physical limits which prevent $\theta$ to
become too large, we have performed MD
simulations in order to deepen our study and check our previous
conjectures. We simulate, by means of an event-driven algorithm a
system equivalent to the experimental ones with $N$ {\em smooth}
disks moving in a plane with DLW, gravity acting along $-y$, a
thermostat (at temperature $v_0^2$) at the bottom wall and a topmost wall which has the same
inelasticity as the lateral walls, see~\cite{sm} for details. The
collisions between disks are treated with a normal restitution
coefficient $\alpha$, while the disk-wall interactions occur with a
normal restitution coefficient $\alpha_w$ (no tangential dissipation
is taken into account). We have also checked that friction with the
plate and rotations/roughness of disks do not change in a significant
way the outcomes of the simulation, confirming that they do not play a
relevant role also in the experiment.
In Fig.~\ref{fig:sim}A we display an example of results from MD with
parameters similar to the experimental setup.  The comparison is very
good in the shape of convection cells as well as in the density and
temperature fields. Performing simulations with many values of all the
parameters we confirmed that convection is always present, excluding
the very dilute Knudsen-gas regime, where the mean free path is larger
than the system lateral size (exactly like in the extremely dilute
experiments). Fig.~\ref{fig:sim}B shows that as soon as the lateral
walls become elastic ($\alpha_w=1$) the convective cells disappear and
the density/temperature fields become homogeneous along $x$. Again,
this confirms the nature of the convective phenomena that we are
observing, as well as the fact that BBD-TC is not acting in our system because of low gravity~\cite{meerson1,meerson2} .
Further evidence comes from simulations at increasing system's width
$L_x$, while keeping fixed all the other parameters, as well as
$N/L_x$. An example of the results is displayed in
Fig.~\ref{fig:width}A. When the width is increased, it appears that
the cells have an intrinsic horizontal size $L_c$ because when $L_x>2
L_c$ a {\em convection-free space} emerges in between. In fact, this
region can be identified with the bulk of the fluid, the existence of
this region being consistent with predictions from theories without
DLW at low gravity~\cite{meerson1,meerson2}. On the contrary, when
$L_x<2L_c$ (as in our experiment), the two cells squeeze in the
available space. Figures~\ref{fig:width}B-E show the behavior of the
width of the convection cell and of the intensity of convection
(see~\cite{sm} for definitions) as a function of $\alpha_w$ and of
gravity $g_{eff}$. Two major observations emerge: 1) the intensity of
convection decreases linearly with $\alpha_w$, in a way similar to the
observation of~\cite{windows13}; 2) the width of the convective cell
decreases when $g_{eff}$ increases. When $g_{eff} \sim g$ convection
is barely visible: this can explain the results and the phase diagram
described in~\cite{lohse07}. In~\cite{sm} we also show a few more
results from MD where density is increased up to a point where BBD-TC
also appears, independently of $\alpha_w$.

The simulations allowed us to verify also cases with DLW
($\alpha_w<1$) but with $\alpha=1$: the results appear identical to
the results with $\alpha<1$, indicating that the elastic limit is
smooth and that DLW are sufficient to create not only the horizontal
gradient but also the vertical ones and drive the system into DLW-TC
regime. An approximate estimate of the width of the convection zone
from the Boussinesq equation with dissipative lateral
walls~\footnote{N. Brilliantov, private communication.}  yields a
dependence on $\alpha_w$ and $g_{eff}$ which is in fair qualitative
agreement with our observations. The analysis confirms that buoyancy
is involved in the DLW-TC mechanism.


In conclusion we have demonstrated the existence, for granular gases,
of a convective phase induced only by DLW, which becomes important
under low gravity conditions. Coupled horizontal and vertical
gradients, of both temperature and density, distinguish the TC studied
here from that analyzed in previous theories~\cite{meerson1,meerson2}
and observed in some experiments~\cite{lohse07,lohse10}, where
horizontal gradients are absent or irrelevant.  The DLW-TC mechanism
resembles secondary flows dominated by boundary layer effects and
horizontal gradients, such as the tea-leaves
paradox~\cite{e26}. Further theoretical investigation is needed to
provide more quantitative predictions from hydrodynamics.  New
experiments in microgravity are also awaited, as well as possible
geophysical applications in low gravity planets, moons and
asteroids. We remark that, according to Eqs.~\eqref{2d_balance},
convection appears whatever is the magnitude (i.e. no threshold for
convection onset) and origin of the horizontal thermal gradient,
suggesting a broad validity also outside of the realm of
granular fluids.

\begin{acknowledgments}
We wish to thank MD Deen and R. Scaccia for help with the experimental setup,
A. Lasanta Becerra for useful discussions and suggestions and
A. Santos for a careful reading of the
manuscript. F. V. R. acknowledges support from grant ”Salvador
Madariaga” No. PRX14/00137 and project No. FIS2013-42840-P (both from
Spanish Government) and through project No. GR15104 (Regional
Government of Extremadura, Spain, partially financed by the European
Regional Development Fund, ERDF).
\end{acknowledgments} 


\newpage

{\bf SM: SUPPLEMENTAL MATERIAL}

\section{Details of the experimental setup}

Our experimental setup consists in a granular partial monolayer moving
on an inclined plane and fluidized by the action of a vibrating
piston. We refer to Fig. 1 of the Letter for a visual sketch of the setup.

$N$ steels spheres (diameter $d=1~\mathrm{mm}$, mass $m=4.3\; $ mg)
move, rolling and sliding, on top of a plate made of aluminum alloy
which is inclined by an angle $\theta$ with respect to the
horizontal. The part of the plate where spheres move upon is a
rectangular area of dimensions $L_x \times L_y$, with
$L_x=175~\mathrm{mm}$ and $L_y=600~\mathrm{mm}$. This area is
delimited by four ``walls'': three of them (we call them top, left and
right) are at rest, while the fourth (the ``bottom'') is a vibrating
piston. The three walls are made of polycarbonate while the piston is
a Plexiglas\textsuperscript{\textregistered} slab. The piston vibrates
with an almost harmonic law $y_p(t) \approx A sin(2\pi f t)$ thanks to
a crankshaft driven by a dc electric motor.  Parallel to the aluminum
plate, at a distance smaller than $2$ spheres' diameters,
$L_z=1.5~\mathrm{mm}$, we have placed a transparent ``roof'' made of
thin glass. Particular care has been taken in order to discharge
electrostatics, by planting many copper cables in the interior of the
plate (from below), all of them connected to a large metal mass (the
optical table where the experiment is mounted on). The use of glass
for the top cover has improved (with respect to
Plexiglas\textsuperscript{\textregistered}) the removal of
electrostatic charges.

A high speed camera (Photron Mini Ux50) records images (parallel to the
plate) of width $L_x$ and height $1.2 \times L_x < L_y$ with minimum
ordinate corresponding to the maximum position of the piston,
i.e. excluding the topmost region of the system which is indeed very
dilute. The image acquisition follows a pairwise protocol: two close
frames (at a distance of $0.002$ s) are recorded, then $2$ seconds are
awaited and the cycle is repeated. The two close frames allow to
determine the position and velocity vectors ${\bf x}_i(t),{\bf
  v}_i(t)$ of the spheres $i \in [1,N]$ (of course it is not
guaranteed that in each frame all particles are recognized and
tracked). The choice of the time intervals is optimized in order to
fulfill the following criteria: the two frames in a pair must be close
enough to catch a ballistic (i.e. non-colliding) trajectory, most of
the time; the two frames in a pair cannot be too close otherwise image
noise cannot be distinguished from real movement; the interval between
two pairs of frames must be large enough to improve statistical
independence. The methods for locating particles and reconstruct their
velocities have been described in previous
references~\cite{puglisi12}.

\section{Study of single particle dynamics in the experiment (evaluation of particle-plate friction effects)}

A preliminary study of the single particle (non-interacting)
trajectories in the experiment has confirmed that the motion in the $xy$ plane of the
center of mass of a particle obeys the following equation:
\begin{equation} \label{dyn}
\ddot{\bf r}(t) = -g_{eff} \hat{\bf y} - \gamma \dot{\bf r}(t) - \mu \frac{\dot {\bf r}(t)}{|\dot {\bf r}(t)|}
\end{equation}
with effective gravity $g_{eff} \approx 5/7 \sin(\theta) g$ (where $g$
is the normal acceleration of gravity on Earth), viscous coefficient
$\gamma= (1.2 \pm 0.4) s^{-1}$ and Coulomb-like (sliding) friction
coefficient $\mu=(80 \pm 20)~mm/s^2$. We have also observed that in
the rising part of each trajectory ($\dot{y} >0$), which usually has a
larger initial velocity (after a collision with the energetic piston),
viscosity is lower and Coulomb friction is larger, while the opposite
occurs in the falling part. This could be due to different modes of
motion, e.g. sliding versus rolling. Since it is impossible, with the
present setup, to retrieve information about the rotational motion of
each sphere, we cannot investigate experimentally this issue. We
notice however that the factor $\approx 5/7$ in $g_{eff}$ indicates
that the trajectories are dominated by pure rolling.

In the simulations, we have verified that using the real values of
$\mu$ and $\gamma$ or putting $\mu=\gamma=0$ does not
change the qualitative picture discussed in the
Letter. Quantitatively, the hydrodynamic fields are affected by those
values in a way which we do not consider significant.

\section{Definition of quantitative measurements in experiments and simulations}

The system (real or simulated) after a very short transient is in a
stationary state and therefore, once the 2D-vectorial positions and
velocities of the particles ${\bf x}_i(t),{\bf v}_i(t)$ are known for
$n_f$ frames at times $t \in [1,n_f]$, the coarse-grained
``hydrodynamic'' fields can be obtained by the following definitions:
\begin{align}
n({\bf x}_k) &= \frac{1}{n_f } \frac{1}{||B_k||}\sum_{t=1}^{n_f} \sum_{i: {\bf x}_i(t) \in B_k}1 \\
{\bf u}({\bf x}_k) &= \frac{1}{n_f } \frac{1}{n({\bf x}_k) ||B_k||}\sum_{t=1}^{n_f}\sum_{i: {\bf x}_i(t) \in B_k} {\bf v}_i(t) \\
T({\bf x}_k) &= \frac{1}{n_f }\frac{1}{n({\bf x}_k) ||B_k||}\sum_{t=1}^{n_f} \sum_{i: {\bf x}_i(t) \in B_k} \frac{|{\bf v}_i(t)- {\bf u}({\bf x}_k,t)|^2}{2},
\end{align}
where ${\bf x}_k$ is the coordinate of the $k$-th point of the mesh
and $B_k$ is the cell of the mesh centered at ${\bf x}_k$, whose area
we call $||B_k||$. In Fig. 2 (experimental fields) we have used a $40
\times 40$ mesh, i.e. $k \in [1,1600]$. In Fig. 3 (simulations) we
have used a $20 \times 20$ mesh, while in Fig. 4A (simulation with a
large system) we used a $80 \times 20$ mesh. In both experiments and
simulations we used $n_f = 1090$.  Temperature and density fields are
shown, in Figs. 2 and 3, through a surface interpolation procedure
called ``pm3d map'' in the {\em gnuplot} software.

In Fig. 4B-E we have also presented two quantitative characterizations
of the convective cells, called ``convective width'' and ``convection
intensity''. Both quantities are obtained by the following procedure:
1) the center of the convective cell is individuated in the plot of
${\bf u}({\bf x}_k)$, in particular its ordinate $y_c$. 2) the
vertical component of the average flow at that ordinate,
$u_y(x,y=y_c)$ always presents the following oscillatory behavior  
: at $x \approx
-L_x/2$ it takes its (negative) minimum value $u_y^{min}$, at $x=x_c$
it goes through zero, then at some larger coordinate it reaches a
maximum value $u_y^{max}$ and then (if $L_x$ is large enough to leave
space for the bulk non-convective region) it quickly goes to zero,
touching it at a point $x_0$ and finally fluctuating around zero for $x_0<x<0$; this pattern, of course, repeats
specularly in the right half of the system ($0<x<L_x/2$); 3) the ``convection
intensity'' is calculated as $u_y^{max}-u_y^{min}$ (in Fig. 4C and 4E
it is plotted after being rescaled by $v_0$); 4) the ``convective
width'' is defined, simply, as the final vanishing point $x_0$ of the
vertical field. In the experiment it is quite difficult to repeat the procedure
because the total width $L_x$ constrains the two cells, i.e. there is
not a point where $u_y(x,y=y_c)$ goes to zero.

\section{Details of the Molecular Dynamics simulations}

We have implemented a Molecular Dynamics simulation with $N$ smooth
disks of diameter $d$ moving in a rectangular box of dimensions $L_x
\times L_y$. The disks follow Eq.~\eqref{dyn} when they are not
overlapping. When two disks touch, a momentum-conserving inelastic
instantaneous collision occurs with restitution coefficient $\alpha
\in [0,1]$ (where the elastic case is given by $\alpha=1$)~\cite{poeschel}. The same
kind of collision occurs when a disk touches the lateral walls at
$x=\pm L_x/2$ and the top wall at $y=L_y$, with the difference that
a wall is treated as an infinite mass particle. A collision with the
bottom wall ($y=0$) conserves horizontal velocity and totally refreshes
the vertical one: the new $v_y$ of the disk is extracted randomly with
a distribution $P(v_y)=(v_y/v_0^2) \exp(-v_y^2/(2v_0^2))$ restricted of
course to $v_y>0$, which defines the average energy of the thermostat
$v_0^2$.

As mentioned, for all the choices of physical parameters ($N$,
$g_{eff}$, $v_0$ etc.) we have performed simulations both with
realistic values of $\gamma$ and $\mu$ and with $\gamma=\mu=0$, and we
have not found substantial differences. In the Letter we have shown
the results with the realistic $\gamma$ and $\mu$.

\section{Evaluation of mean free paths and Knudsen number in the experiment and in the simulations}

Since a systematic theoretical framework for the phenomena discussed
in our Letter is still lacking, we have decided to give most of the parameters
of both experiments and simulations in raw physical units. In
the Figures of the Letter we have used adimensional units (e.g. length
is rescaled by diameter, temperature field is rescaled by the piston
average energy, etc.)  for simplicity, without claiming that those
reference scales have any theoretical value.

In this section we discuss possible relevant scales, similar to the one used in a previous work~\cite{VU09}, in view of a
theoretical approach based upon granular gas-hydrodynamics~\cite{poeschel,puglio15}.

\subsection{The Knudsen number reference scale}

In standard kinetic theory of non-uniform gases, a usual choice of
length and time reference units is the mean free path and collision
frequency~\cite{CC70}. The choice of these units emerges naturally from the
collisional frequency prefactors that appear in the collisional
integrals associated to the transport coefficients of the gas (either
molecular gas or granular gas). At the level of the average fields
representation, it is also useful this choice of units since it
straightforwardly yields the spatial gradients in terms of a reference
Knudsen number.

Mean free path and collision frequency can be measured in a generic reference point at local density $n_r$ and local temperature $T_r$ \cite{CC70} using the following formula valid in dimension $2D$:
\begin{equation}
\lambda_r=(\sqrt{2\pi}n_r d)^{-1}\: ,
\label{lambda2D}
\end{equation} 
\begin{equation}
\nu_r=\sqrt{\frac{T_r}{m}}\frac{n_rd}{\sqrt{\pi}}\: .
\label{nu2D}
\end{equation}
As we said, when $\nu_r$, and $\lambda_r$ are chosen as time and space
reference units, complemented with mass particle $m$ for mass, the
spatial gradients $\nabla^{(n)}$ from the balance equations in this
representation are of the order $\mathrm{Kn}^{(n)}$ which is specially
useful since kinetic theory of non-uniform gases is usually developed
as a perturbative theory where the distribution function is developed
in powers of $\mathrm{Kn}$ \cite{BDKS98}. The choice of the reference
point where granular temperature $T_r$ and density $n_r$ are measured
is a subtle question and it may be important to make an optimal choice
\cite{BCC94}.  In the presence of a gravitational field like in our
system we may set our reference point at the bottom piston, since - at
least at the theoretical level - temperature and density are constant
in its proximity; i.e. we can use $T_r=T(x,y=0)$ and $n_r=n(x,y=0)$. A
different choice could be more involved since at any other distance
$y$ of the piston, temperature and density will also be a function of
$x$ due to the dissipative lateral walls (DLW). Due to the complex
(non-hydrodynamic) boundary layer problem, each experiment yields
different values of $T_r=T(x,y=0)$ and $n_r=n(x,y=0)$ and therefore
the scaling cannot be anticipated, a real measure of $T_r$ and $n_r$
is necessary. For instance, the rescaled dimensions $\hat L_x\equiv
L_x/\lambda_r$, $\hat L_y\equiv L_y/\lambda_r$, which can be indicated
as first estimates of the Knudsen number, are different for each
experiment, even if the real size is constant. Some examples of
adimensional quantities in the real setup are shown in
table~\ref{tabla}.

\subsection{The gravity reference scale}

A complementary dimensionless representation may be obtained by referring
the system size to the length $l_g=v_r^2/g$, with $v_r$ a
reference thermal velocity. The thermal velocity is also a
non-homogeneous quantity but we can take as a reference its value next
to the piston, i.e. for instance $v_r=\sqrt{2 T_r/m}$. Of course it is important that
the lengthscale induced by gravity is not smaller than the mean free
path, i.e. we need a small $\hat g\equiv g\lambda_r/v_r^2$. This can be checked in table~\ref{tabla}.


In table~\ref{tabla} we give also examples of rescaled length through the gravity reference scale.
As we can see in the corresponding values of $L_x/(l_g)=\hat
L_x\hat g$ and $L_y/(l_g)=\hat L_y\hat g$, they are not too large
for all values of $N$, specially for low $N$. 


\hspace{0.5mm}

\begin{table}
\begin{center}
\begin{tabular}{ |l|l|l|l|l|l| }
\hline
$N$ & $\hat L_x$ & $\hat L_y$ & $\hat g$ &  $\hat L_x\hat g$ &  $\hat L_y\hat g$ \\ 
  \hline\hline
    100  & 0.96  & 0.96  & 0.211 & 0.202  & 0.693 \\ \hline
    200  & 2.19  & 2.20  & 0.127 & 0.277  & 0.950 \\ \hline
    300  & 4.76  & 4.78  & 0.086 & 0.413  & 1.416 \\ \hline
    500  & 14.61  & 14.64  & 0.054 & 0.790  & 2.709 \\ \hline
    700  & 32.72  & 32.80  & 0.047 & 1.549  & 5.312 \\ \hline
    1000  & 76.79  & 76.98  & 0.023 & 1.755 & 6.018 \\ \hline
  \hline
\end{tabular}
\end{center}
\caption{Rescaled dimensions and gravity in a series of real
  experiments with $f=45~\mathrm{Hz}$, $A=1.85~\mathrm{mm}$ and
  $g_{eff}=0.016~g$ and different values of $N$.}\label{tabla}
\end{table}

\section{Transition to buoyancy-driven thermal convection}

At variance with the DLW-induced convection studied in the Letter,
which is always present if $\alpha_w<1$, buoyancy-driven thermal
convection in granular gases occurs in a limited region of parameters:
linear stability of hydrodynamics without lateral boundaries (see
Ref. [22] in the Letter) predicts that, at constant gravity and base
temperature, it is triggered by an increase of the effective
inelasticity $R \propto (1-\alpha) n^2$ where $n$ is the average
number density $N/(L_x L_y)$. Moreover, the typical size of the
convective cells is predicted to increase with gravity (see Fig. 5 of
Ref. [22]).

We have tried to see how buoyancy-driven convection appears in our MD
simulations at low gravity. We have chosen a large system (identical
to Fig. 4A in the Letter), in order to put in evidence the bulk region
which is not affected by boundaries and to enhance the possibility to
accommodate bulk-convective cells. Two different situations have been
analyzed, one with inelastic lateral boundaries and another one with
elastic walls. In both cases, the increase of $N$ (and consequently of
$R$) determines an appearance of buoyancy-driven thermal convection,
which is characterized by {\em many} convective cells spanning the
whole width of the system. The following Figure 1 explains the
situation:

\begin{figure}[h]
\includegraphics[width=5cm,clip=true,angle=-90]{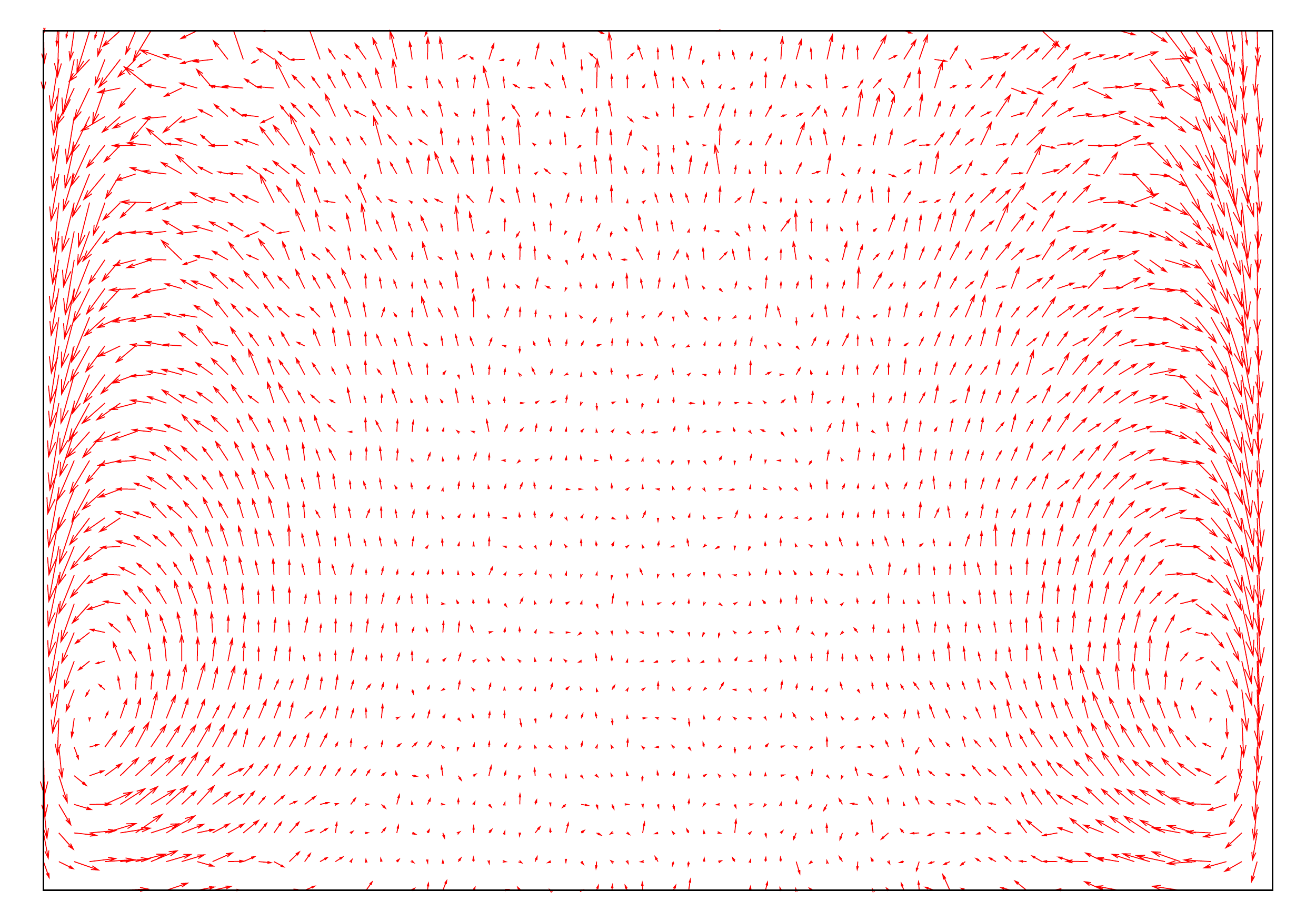}
\includegraphics[width=5cm,clip=true,angle=-90]{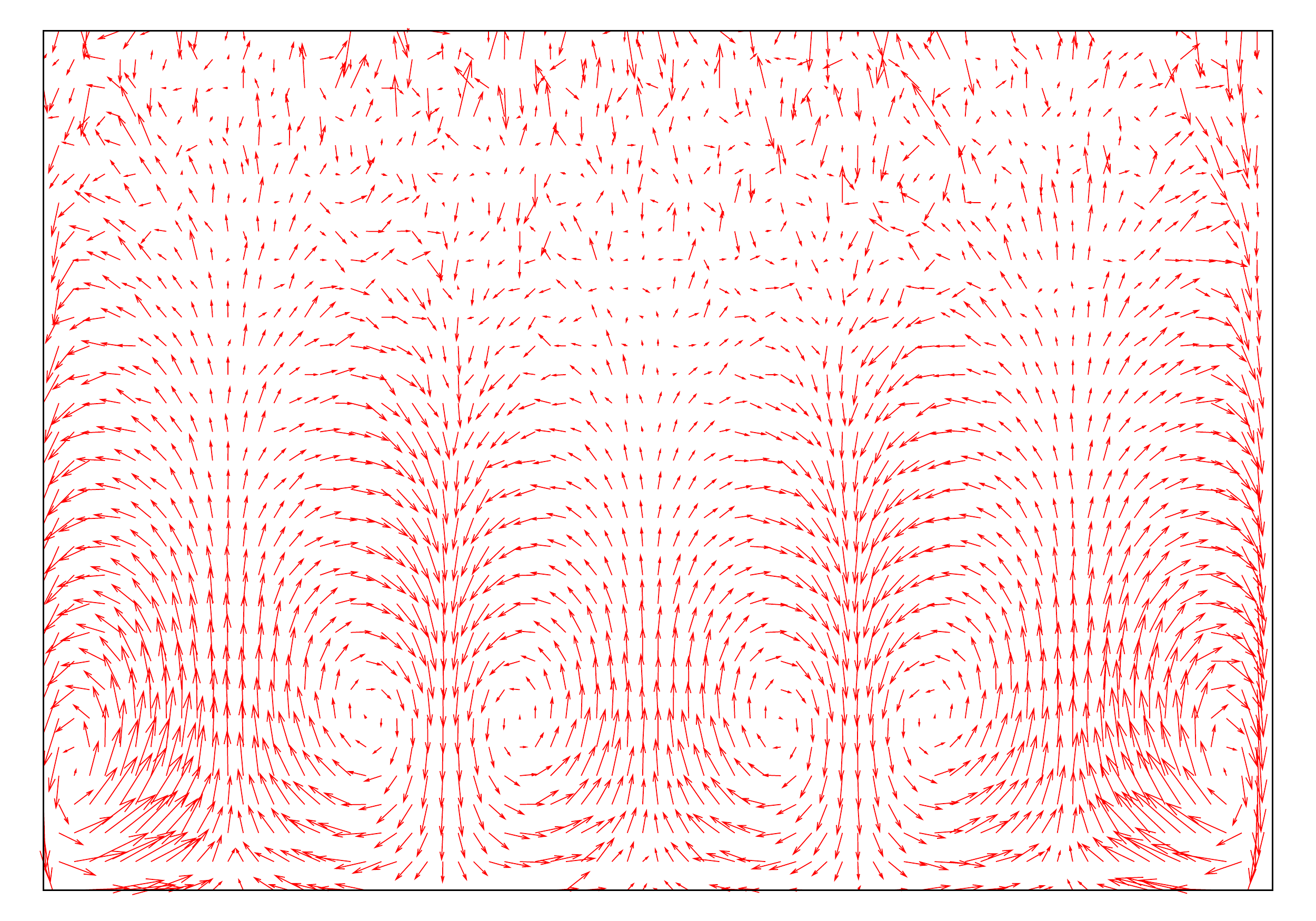}\\
\includegraphics[width=5cm,clip=true,angle=-90]{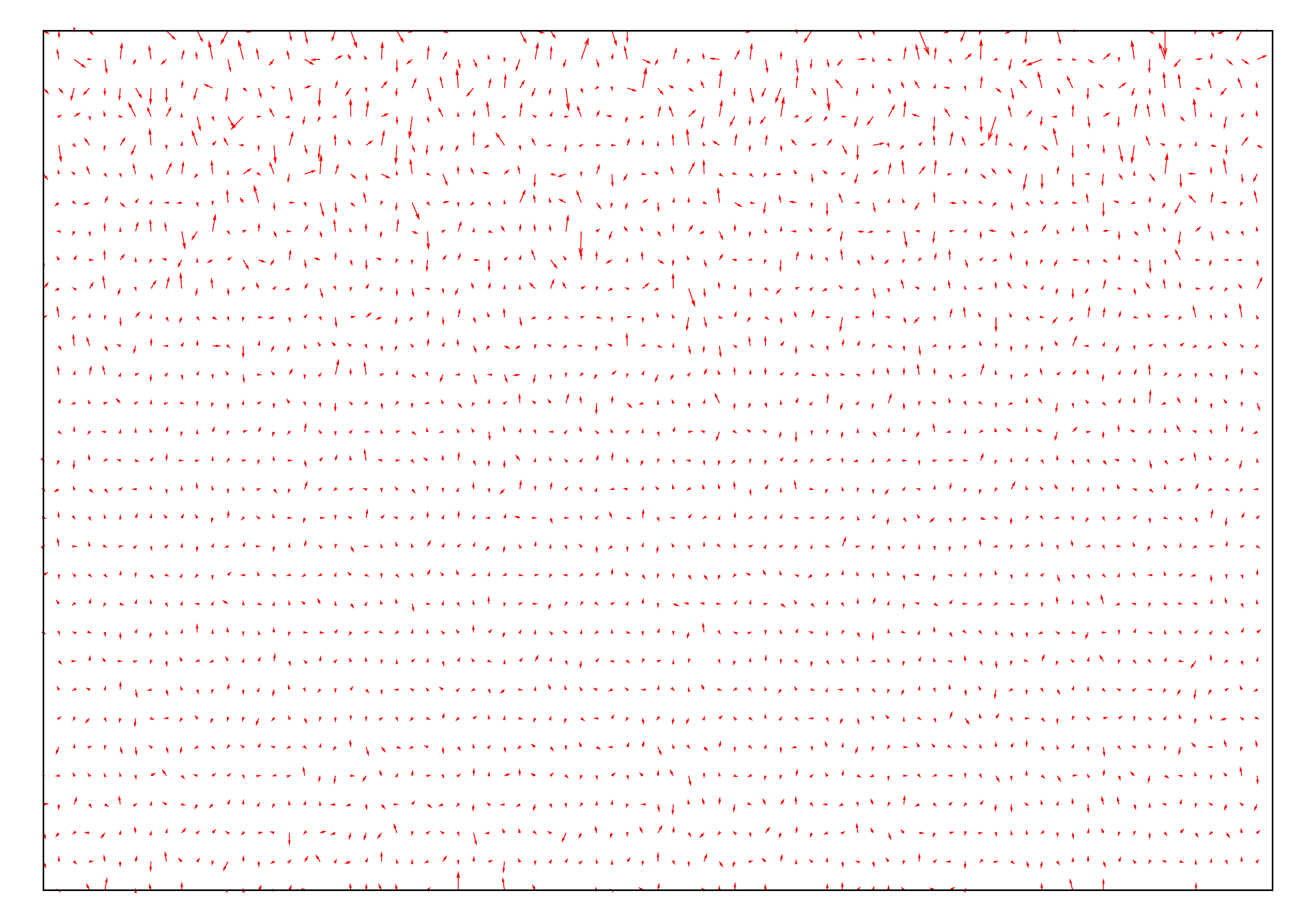}
\includegraphics[width=5cm,clip=true,angle=-90]{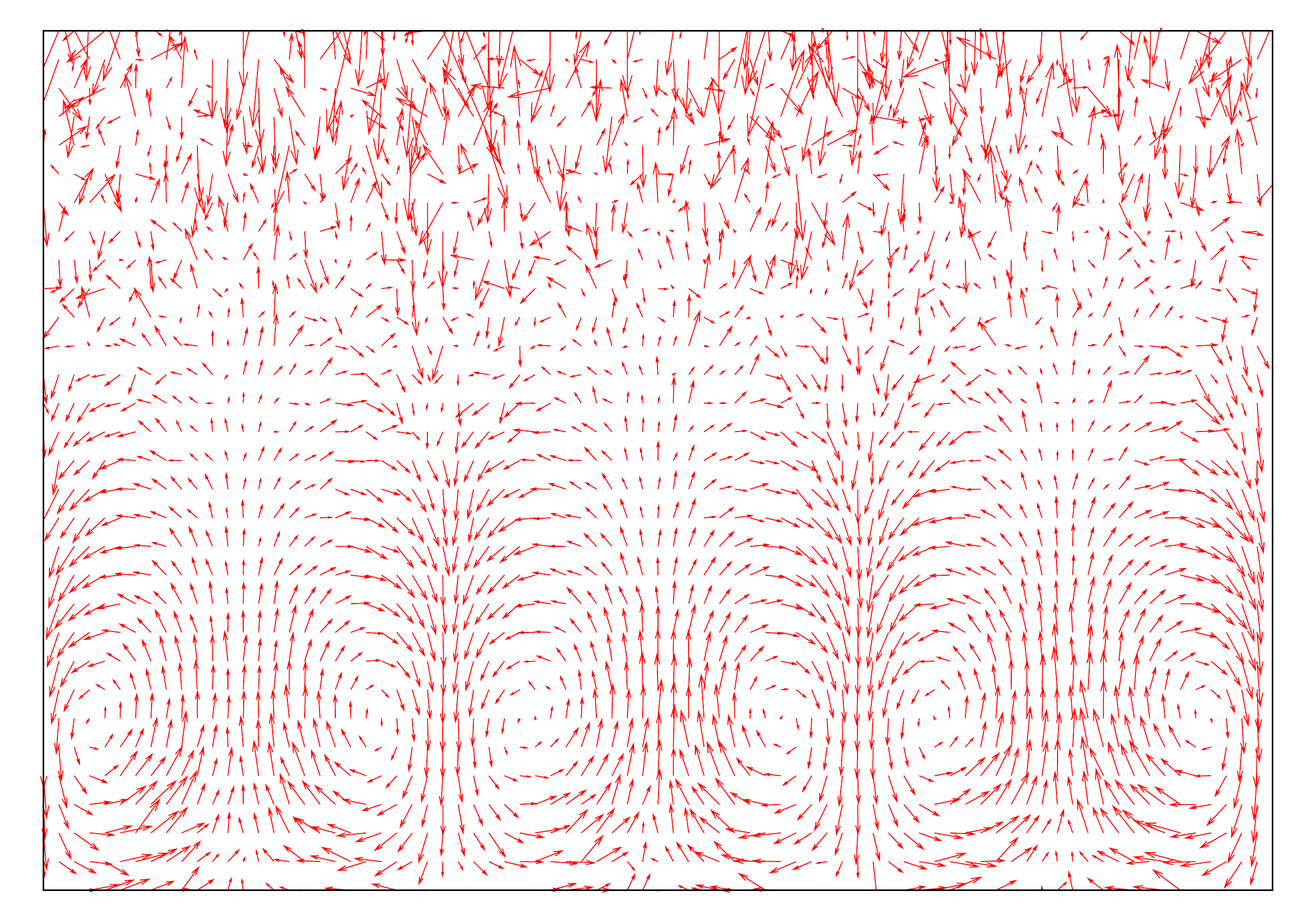}\\
\caption{Top left: $N=2800$, $r_w=0.3$ (Bulk convection: no. Dlw convection: yes.). Top right: $N=6000$, $r_w=0.3$ (Bulk convection: yes. Dlw convection: superimposed.)
Bottom left: $N=2800$, $r_w=1$ (Bulk convection: no. Dlw convection: no.). Bottom right: $N=6000$, $r_w=1$ (Bulk convection: yes. Dlw convection: no.)}
\end{figure}

Below (Figure 2) we report also the plot of the density (packing fraction) field, which is useful to get
an idea of the relevance of clustering phenomena: those are expected
to play a role when inelasticity or density are large. It is clear
that, in the absence of buoyancy-driven convection, there is no
clustering (apart from near inelastic walls). The appearance of bulk
convection also induces inhomogeneities of the density field, as usual
(see for instance Ref. [28] of the Letter).

\begin{figure}[h]
\includegraphics[width=5cm,clip=true,angle=-90]{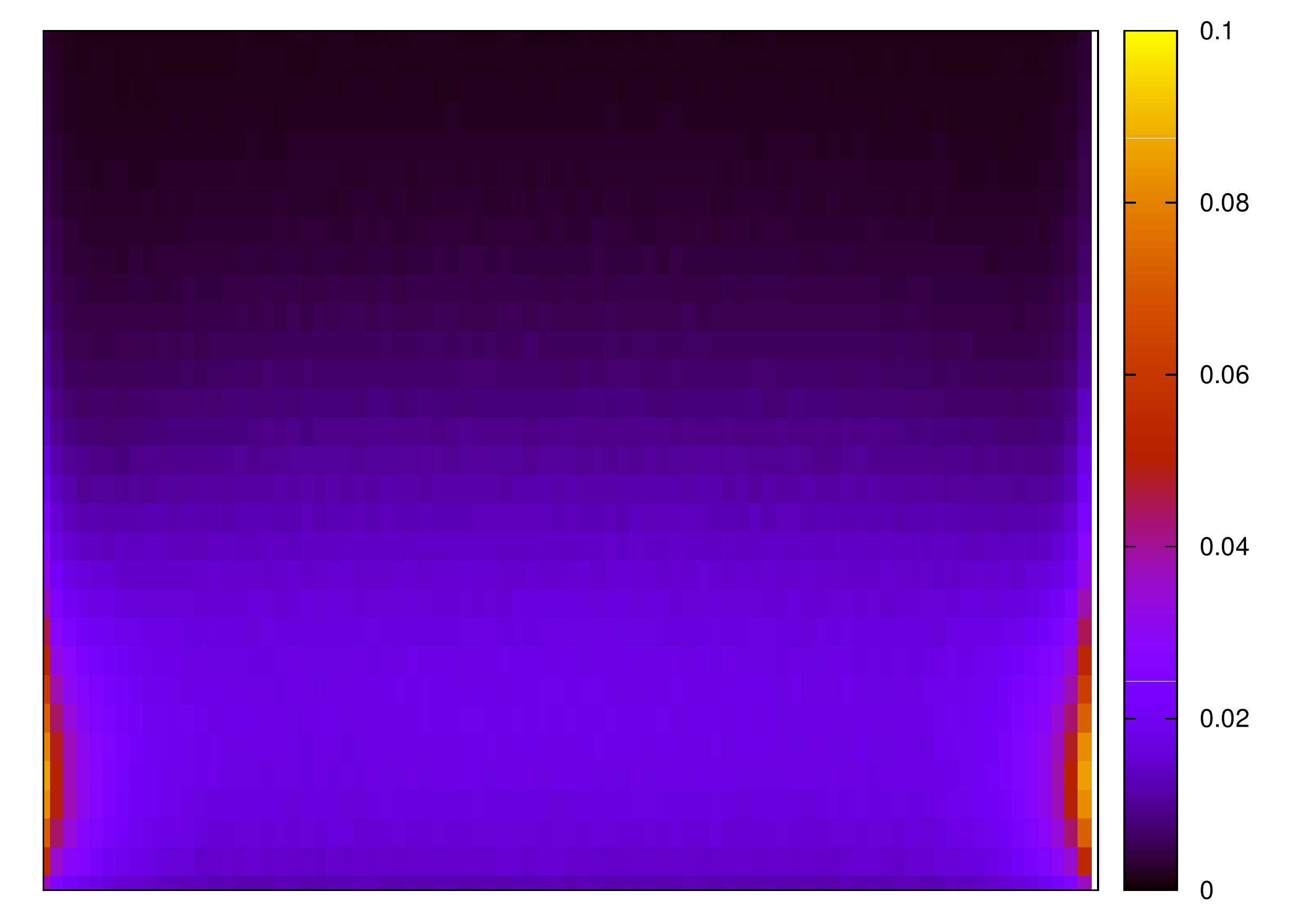}
\includegraphics[width=5cm,clip=true,angle=-90]{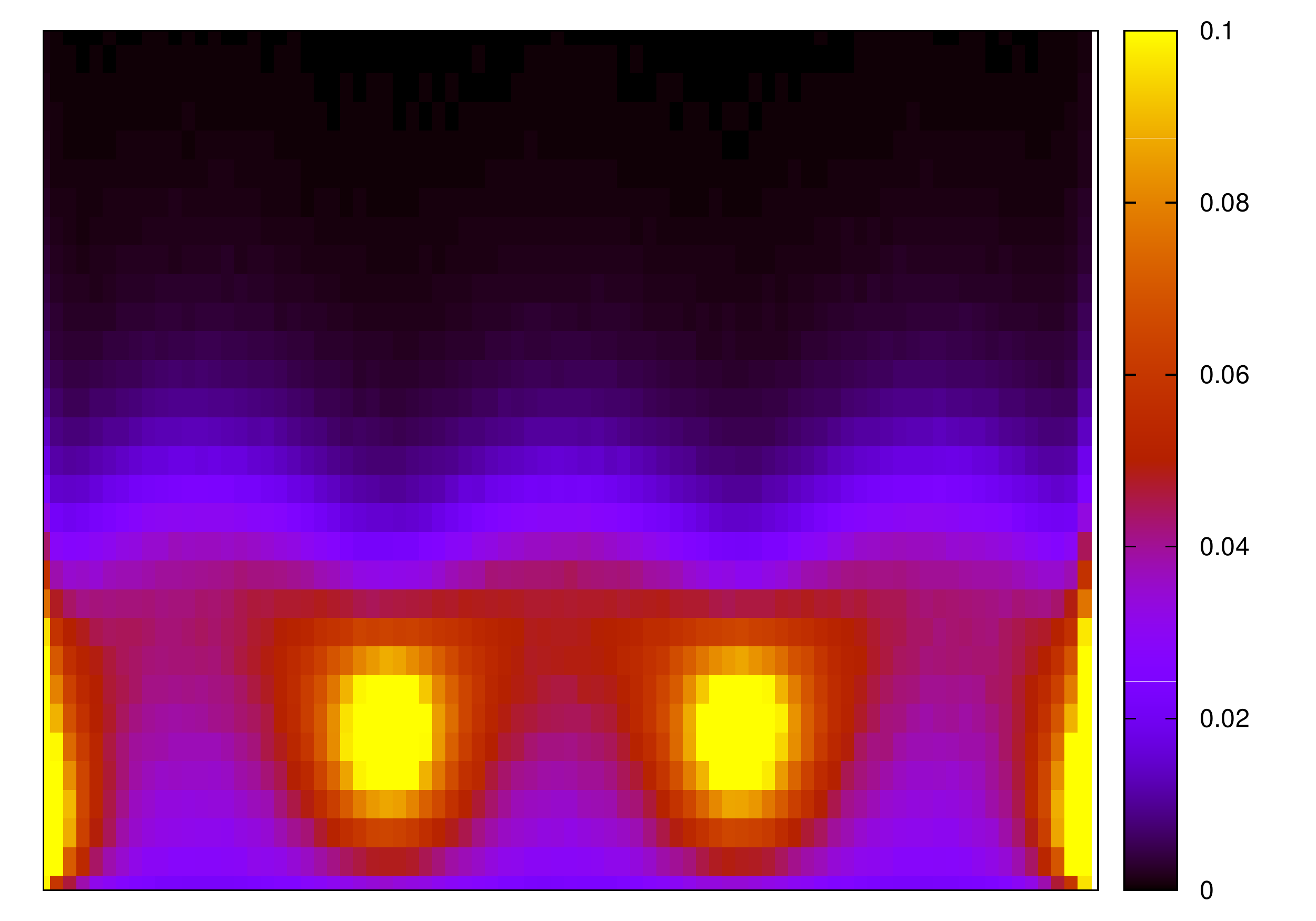}\\
\includegraphics[width=5cm,clip=true,angle=-90]{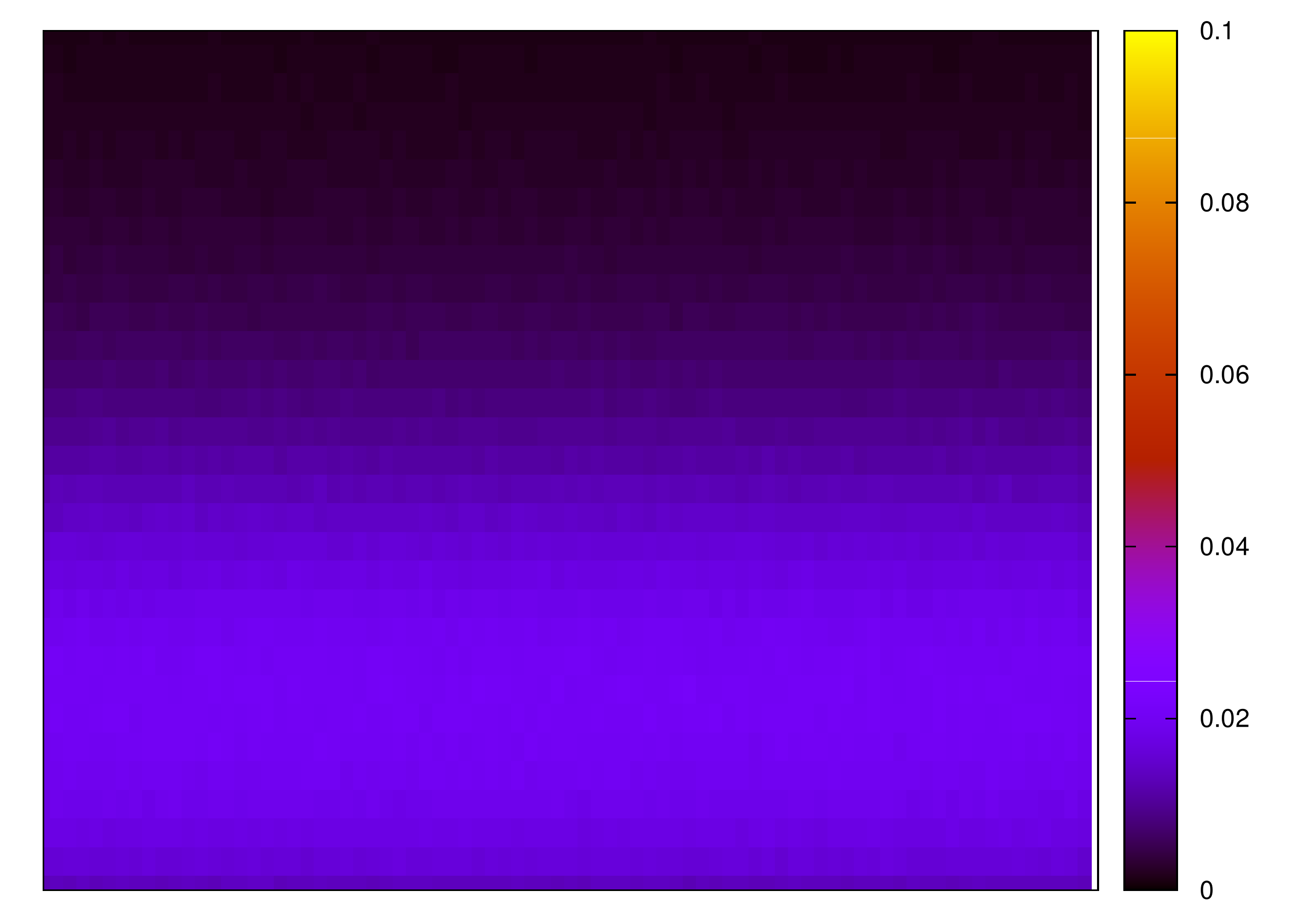}
\includegraphics[width=5cm,clip=true,angle=-90]{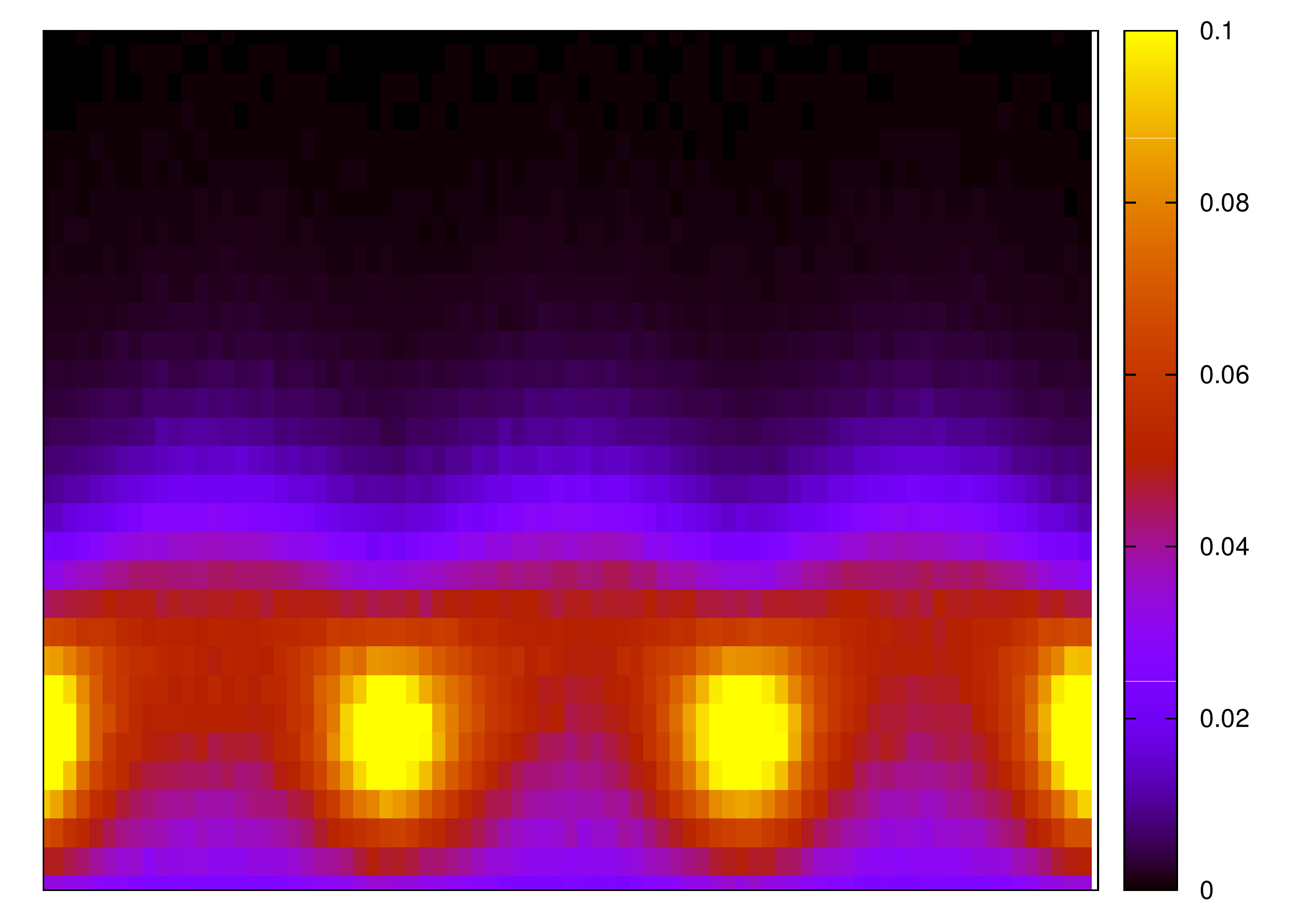}\\
\caption{Local packing fraction field. Top left: $N=2800$, $r_w=0.3$ (Bulk convection: no. Dlw convection: yes.). Top right: $N=6000$, $r_w=0.3$ (Bulk convection: yes. Dlw convection: superimposed.)
Bottom left: $N=2800$, $r_w=1$ (Bulk convection: no. Dlw convection: no.). Bottom right: $N=6000$, $r_w=1$ (Bulk convection: yes. Dlw convection: no.)}
\end{figure}

\bibliography{biblio_merged}

\end{document}